\documentclass[preprint,longtitle,3p,twocolumn,times]{elsarticle}

\usepackage{footnote}
\makesavenoteenv{tabular}
\makesavenoteenv{table}
\usepackage{amsmath}
\usepackage[hyphens]{url}
\usepackage{caption,subcaption}

\journal{Astropart. Phys.}
\begin{document}

\newcommand{\pfreqm}{\nu_{e}}
\newcommand{\pfreqradm}{\omega_{e}}
\newcommand{\cfreqm}{\nu}
\newcommand{\freqm}{\nu_{r}}
\newcommand{\freqradm}{\omega_{r}}
\newcommand{\pfreq}{$\pfreqm$}
\newcommand{\pfreqrad}{$\pfreqradm$}
\newcommand{\cfreq}{$\cfreqm$}
\newcommand{\freq}{$\freqm$}
\newcommand{\freqrad}{$\freqradm$}
\newcommand{\pmax}{$P_{\text{max}}\;$}
\newcommand{\fmax}{$F_{\text{max}}\;$}
\newcommand{\fmaxm}{F_{\text{max}}}
\newcommand{\pmaxm}{P_{\text{max}}}
\newcommand{\xmaxm}{X_{\text{max}}}
\newcommand{\xmax}{$\xmaxm$}
\newcommand{\enotm}{E_{0}}
\newcommand{\enot}{$\enotm$}
\newcommand{\gcmsqm}{\text{g/cm}^{2}}
\newcommand{\gcmsq}{$\gcmsqm$}
\newcommand{\nmaxm}{N_{\text{max}}}
\newcommand{\nmax}{$\nmaxm$}
\newcommand{\dtm}{\delta t}
\newcommand{\dt}{$\dtm$}
\newcommand{\ptm}{P_{T}}
\newcommand{\pt}{$\ptm$}
\newcommand{\gtm}{G_{T}}
\newcommand{\gt}{$\gtm$}
\newcommand{\grm}{G_{R}}
\newcommand{\gr}{$\grm$}

\begin{frontmatter}

\author[1]{R.U.~Abbasi}
\author[2]{M.~Abe}
\author[3]{M.~Abou~Bakr~Othman}
\author[1]{T.Abu-Zayyad}
\author[1]{M.~Allen}
\author[1]{R.~Anderson}
\author[4]{R.~Azuma}
\author[1]{E.~Barcikowski}
\author[1]{J.W.~Belz\corref{cor1}}
\cortext[cor1]{Corresponding Author. Tel.: +01 801 585-9620. Addr.: 115 S 1400 E \#201 JFB}
\ead{belz@physics.utah.edu}
\author[1]{D.R.~Bergman}
\author[5,6]{D.~Besson}
\author[1]{S.A.~Blake}
\author[1]{M.~Byrne}
\author[1]{R.~Cady}
\author[7]{M.J.~Chae}
\author[8]{B.G.~Cheon}
\author[9]{J.~Chiba}
\author[10]{M.~Chikawa}
\author[11]{W.R.~Cho}
% ...

\address[1]{High Energy Astrophysics Institute and Department of Physics and Astronomy, University of Utah, Salt Lake City, Utah, USA}
\address[2]{The Graduate School of Science and Engineering, Saitama University, Saitama, Saitama, Japan}
\address[3]{Department of Electrical and Computer Engineering, University of Utah, Salt Lake City, Utah, USA}
\address[4]{Graduate School of Science and Engineering, Tokyo Institute of Technology, Meguro, Tokyo, Japan}
\address[5]{University of Kansas, Lawrence, Kansas, USA}
\address[6]{Moscow Engineering and Physics Institute, Moscow, Russia}
\address[7]{Department of Physics and Institute for the Early Universe, Ewha Womans University, Seodaaemun-gu, Seoul, Korea}
\address[8]{Department of Physics and The Research Institute of Natural Science, Hanyang University, Seongdong-gu, Seoul, Korea}
\address[9]{Department of Physics, Tokyo University of Science, Noda, Chiba, Japan}
\address[10]{Department of Physics, Kinki University, Higashi Osaka, Osaka, Japan}
\address[11]{Department of Physics, Yonsei University, Seodaemun-gu, Seoul, Korea}
% ...

\title{First Upper Limits on the Radar Cross Section of \\ Cosmic-Ray Induced Extensive Air Showers}

\begin{abstract}
TARA (Telescope Array Radar) is a cosmic ray radar detection experiment colocated with Telescope Array, the conventional surface scintillation detector (SD) and fluorescence telescope detector (FD) near Delta, Utah, U.S.A. The TARA detector combines a 40~kW, 54.1~MHz VHF transmitter and high-gain transmitting antenna which broadcasts the radar carrier over the SD array and within the FD field of view, towards a 250~MS/s DAQ receiver. TARA has been collecting data since 2013 with the primary goal of observing the radar signatures of extensive air showers (EAS). Simulations indicate that echoes are expected to be short in duration ($\sim 10~\mu{\rm s}$) and exhibit rapidly changing frequency, with rates on the order 1~MHz/$\mu$s. The EAS radar cross-section (RCS) is currently unknown although it is the subject of over 70~years of speculation. A novel signal search technique is described in which the expected radar echo of a particular air shower is used as a matched filter template and compared to waveforms obtained by triggering the radar DAQ using the Telescope Array fluorescence detector. No evidence for the scattering of radio frequency radiation by EAS is obtained to date. We report the first quantitative RCS upper limits using EAS that triggered the Telescope Array Fluorescence Detector.

\end{abstract}


\end{frontmatter}



\section{Introduction}
\label{sec:intro}
Ultra-high energy cosmic ray (UHECR, primary energy $\enotm > 10^{18}$~eV) research is limited primarily by low flux. There are currently two dominant detection techniques, surface detectors (SD) comprised of plastic scintillator or water Cherenkov detectors, and fluorescence telescope detectors (FD). Conventional detection methods have been successful in mapping out the UHECR spectrum, observing the GZK cutoff~\cite{abbasi2008,abuzayyad2013,abraham2008} and locating a potential ``hotspot'' in CR arrival directions~\cite{TAhotspot}.  However, large SD arrays are expensive to build and maintain, and FD telescopes have limited statistics due to their low duty cycle ($\sim 10\%$). 

Recently proposed alternative detection methods include those using the geo-magnetic synchrotron~\cite{falcke2005} and Askaryan~\cite{Askaryan} effects, both of which require a large instrumentation area. Radar detection has the potential to be being a remote detection technique with 100\% duty cycle. The idea was initially suggested in 1941~\cite{blovell} when reflections from extensive air showers (EAS) were proposed as a possible explanation for anomalous atmospheric radar echoes.

\begin{figure}[h]
\centerline{\includegraphics[trim=0.0cm 0.0cm 0.0cm 0.0cm,clip=true,width=0.5\textwidth,natwidth=5764,natheight=4598]{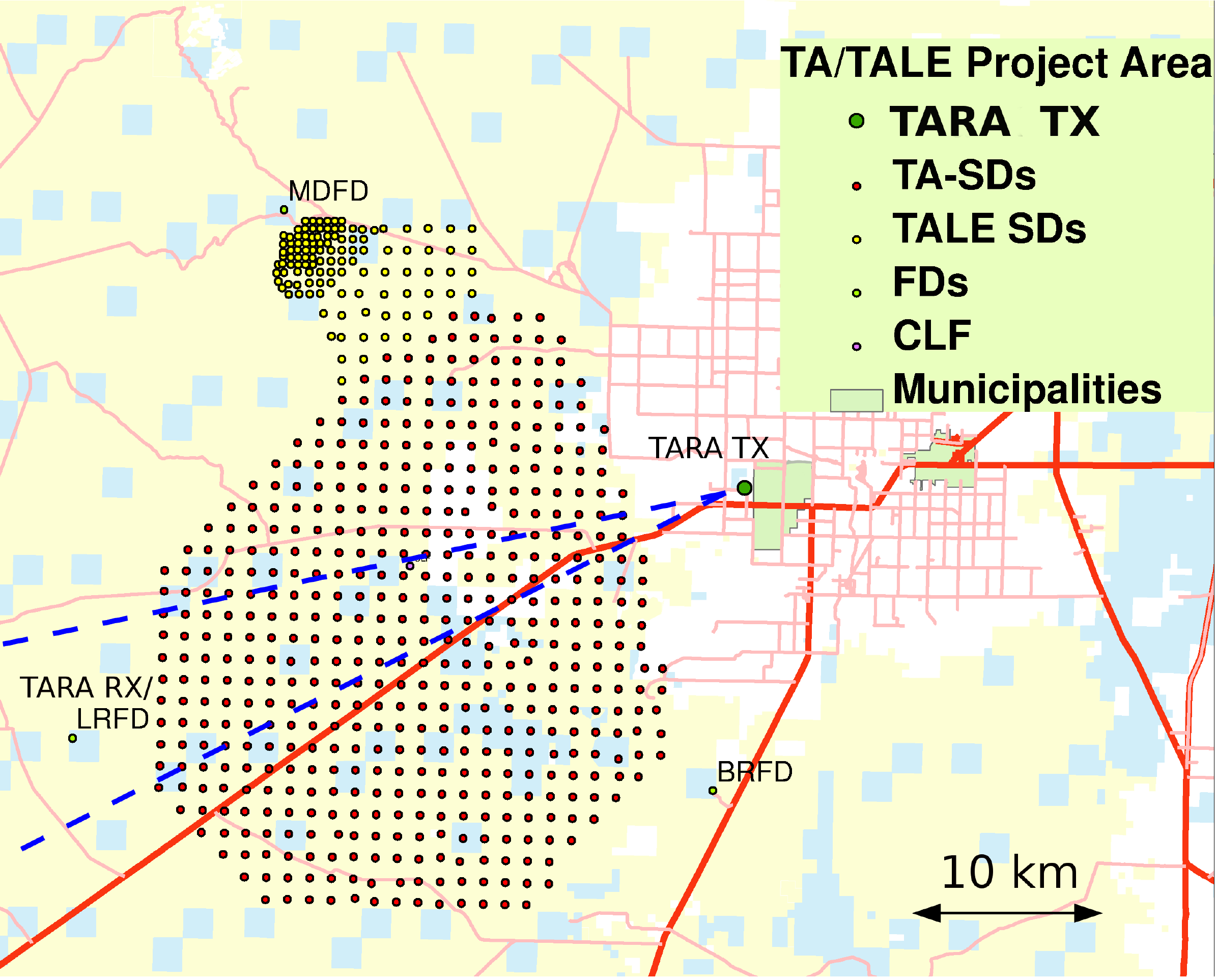}}
{\caption{Map of Telescope Array (TA) detector facilities, both SD and FD, and TARA transmitter and receiver sites. The radar carrier is broadcast from the transmitter site toward the TA Long Ridge Fluorescence Detector (FD). Dashed blue lines indicate the beamwidth at the points 3~dB below the peak gain. This map is copied from~\cite{taranim}.}
\label{fig:tamap}}
\end{figure}

Telescope Array Radar (TARA) seeks to observe radar echoes in coincidence with the Telescope Array (TA) and determine the viability of the radar technique for UHECR detection. TARA is the most advanced CR radar detection experiment to date, improving upon other experiments with several key attributes: 
\begin{itemize}
\item The transmitter is under the direct control of experimenters, and in a radio-quiet area isolated from other radio frequency (RF) sources. The power and radiation pattern are known at all times.
\item Forward power up to 40~kW and gain exceeding 20~dB maximize energy density in the radar field. 
\item Continuous wave (CW) transmission gives 100\% duty cycle, as opposed to pulsed radar. 
\item TARA utilizes a high sample rate DAQ (250~MS/s).
\item TARA is colocated with a large state-of-the-art conventional CR observatory, allowing the radar data stream to be sampled at the arrival times of known cosmic ray events. 
\end{itemize}
Each of these attributes of the TARA detector has been discussed in detail in the literature~\cite{taranim}. A map showing the TA SD array and the location of the TARA transmitter and receiver is shown in Figure~\ref{fig:tamap}.

Section~\ref{sec:scattering} of this paper includes a description of air shower plasmas and possible radio scattering mechanisms. Theoretical and experimental parameters that influence radio scattering are presented and discussed. We justify use of the {\em thin wire model} in a radar echo simulation that predicts echo waveforms, which we will subsequently (Section~\ref{sec:rcscalc}) use in placing limits on the air shower radar cross section (RCS). Sections~\ref{sec:data}~and~\ref{sec:offline} describe TARA data and offline processing techniques. In Section~\ref{sec:sigsearch}, we describe the signal search using simulated waveforms as matched filter (MF) templates in order to maximize sensitivity. Section~\ref{sec:rcscalc} describes the procedure for calculating a scale factor to the RCS model described in Section~\ref{sec:scattering}, the results of which are used in placing the first quantitative upper limit on the EAS radar cross-section (RCS). In Section~\ref{sec:conclusion} we summarize these results and discuss the viability of radar detection of cosmic rays in light of the TARA findings. 


\section{EAS Radio Scattering}
\label{sec:scattering}

We begin with an overview of the issues relevant to RF scattering by EAS, focusing on those which inform the design of the TARA detector and its data analysis.

\subsection{Air Shower Plasmas}
\label{sub:easplasma}

The bi-static radar equation 
\begin{equation}
\label{eq:bi}
P_R = P_T \frac{G_T}{4\pi R^2_T}\,\sigma \,\frac{G_R}{4\pi R^2_R}\frac{\lambda^2}{4\pi}\,,
\end{equation}
is a simple geometrical formula used to calculate received power when the transmitter (TX) and receiver (RX) are at different locations. $P_R$ is received power, $P_T$ is transmitter power, $G_R$ and $G_T$ are the receiver and transmitter station antenna gains, respectively, $\lambda$ is the radar wavelength, $R_T$ is distance between transmitter and target, $R_R$ is distance between target and receiver, and $\sigma$ is the RCS. Most of the parameters in the bi-static radar equation are known \emph{a priori} or can be measured. In the case of EAS, the RCS is currently unknown and, as will be shown, is difficult to calculate directly. 

Scattering occurs from interaction of the radar wave with free electrons liberated from air molecules by passing shower particles. Therefore, the RCS is proportional to the density and area of the plasma body. This merits a discussion of the calculation of ionization density created by EAS. 

Standard EAS parameters include primary particle energy \enot\ (eV), depth \xmax\ (\gcmsq) in the atmosphere where the number of shower particles reaches a maximum \nmax, and the depth of first interaction $X_0$ (\gcmsq). The critical energy $E_{\text{c}}$ (eV) is the energy below which the dominant energy loss mechanism is bremsstrahlung radiation rather than pair production. In air, $E_{\text{c}}$ is 81~MeV~\cite{pdg2014}. When the average particle energy decreases below $E_{\text{c}}$, particle production ceases and the EAS starts to decrease in size.  

Gaisser and Hillas~\cite{GaisHill} have parameterized the average shower longitudinal profile, the number of charged shower products $N(X)$ as a function of depth $X$ in the atmosphere. Nishimura, Kamata, and Greisen (NKG)~\cite{greisen} have created a function describing the number area density as a function of radius, at a specific shower age $s$~\cite{sokolsky}, as a function of $N(X)$. Together, Gaisser-Hillas and NKG describe charged particle density as the shower evolves. Electrons and positrons are the dominant component of charged shower particles~\cite{sokolsky}, outnumbering other species such as muons, to the extent that $N(X)$ can be approximated as the total number of electrons and positrons produced by the shower. This assumption is maintained in this paper. 

Together with the Gaisser-Hillas and NKG functions, atmospheric energy deposition models permit the calculation of ionization yield.  Nerling~et~al.~\cite{nerling} have parameterized $\alpha_{\text{eff}}(X)$, where $\alpha_{\text{eff}}(X)\,N(X) = \frac{dE}{dX}(X)$. Combined with the mean ionization energy in the atmosphere, $I = 33.8$~eV~\cite{rossi1952}, the energy deposit $\frac{dE}{dX}(X)$ from Nerling can be used to estimate the EAS plasma properties. Atmospheric ionization occurs at a greater rate in regions where the shower particle density is high. Therefore, it is expected that the total plasma density follows the shower lateral distribution described by NKG. 

\begin{figure}[h]
\centerline{\includegraphics[trim=0.0in 0.2in 0.5in 0.3in, clip=true, width=0.45\textwidth]{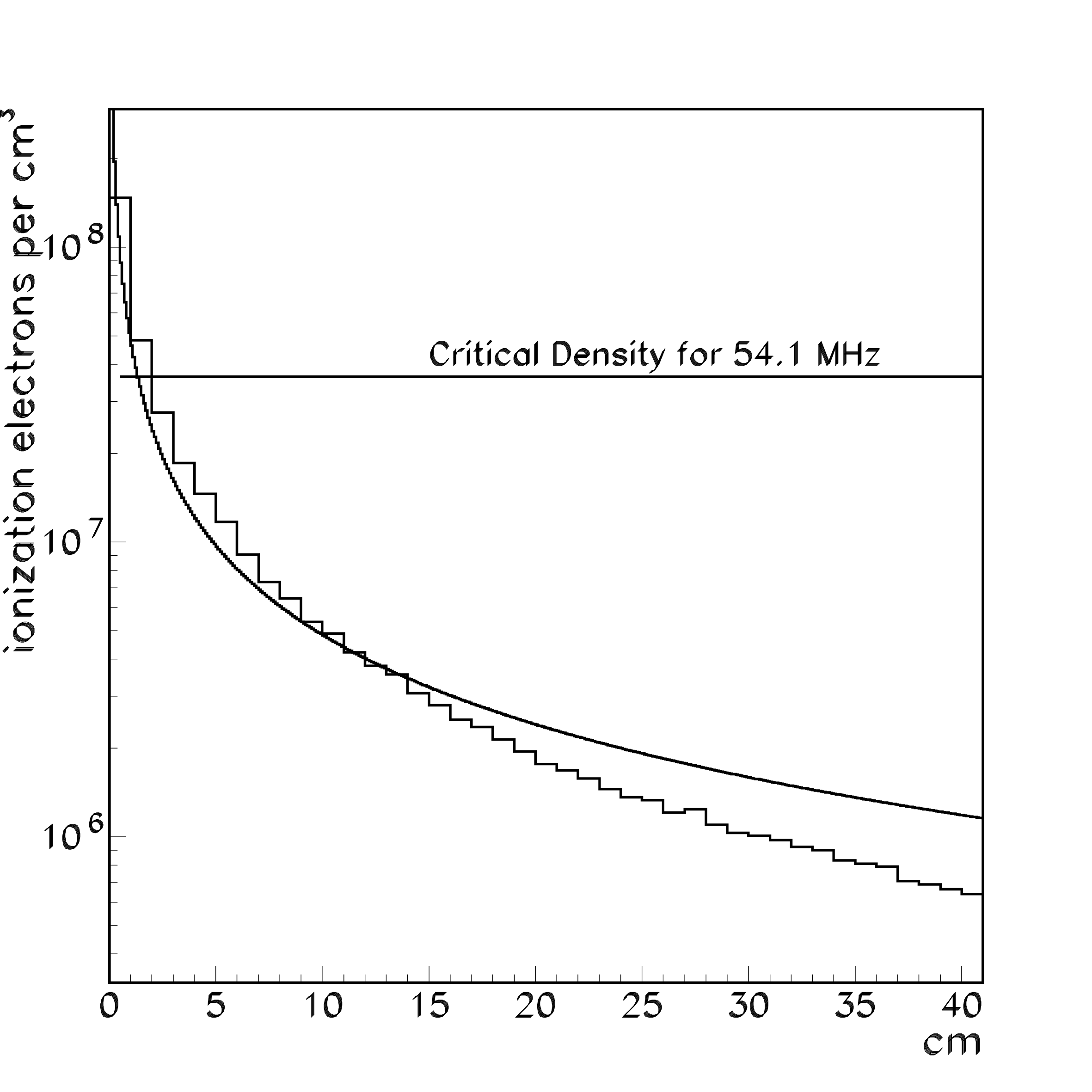}}
{\caption{The comparison of the ionized plasma densities calculated with CORSIKA (histogram) and with Gaisser-Hillas and NKG functions (curve) as a function of radius near \xmax\ for a $10^{19}$~eV vertical shower.} 
\label{fig:odcomp}}
\end{figure}

Figure~\ref{fig:odcomp} shows the ionization density predicted by the NKG distribution for an $\enotm=10^{19}$~eV air shower at \xmax, compared with an unthinned simulated shower produced with the CORSIKA~\cite{heck1998} package using the QGSJETII-03~\cite{ostapchenko2006}) high-energy interaction model. The ionization electron density is calculated for the CORSIKA shower as follows:
\begin{equation}
\label{eq:rnions}
N_{ions} = \frac{E_e \times (1-\frac{1}{e}) \times \rho_{air}}{X_{rad} \times E_{ion}}
\end{equation}
where $E_e$ is the electron or positron energy from CORSIKA, $\rho_{air}$ is the density of air at 1500~m M.S.L, $X_{rad}$ is the electron radiation length in air, and $E_{ion}$ is the mean ionization energy of air. Agreement between the NKG and CORSIKA predictions is good near the critical region close to the shower core, where electron density is highest. 

\subsection{Plasma Scattering}
\label{sub:plasma_scatt}

In the presence of an incident time-varying electric field, a low density plasma will oscillate according to the equation of motion:
\begin{equation}
\label{eq:pfreq_eom}
e\mathbf{E} = m\frac{\partial^2 \mathbf{r}}{\partial t^2}\,,
\end{equation}
where $\mathbf{r}$ is the electron displacement vector. Both $\mathbf{E}$ and $\mathbf{r}$ have the harmonic component $\text{exp}(-i\freqradm t)$, with \freqrad\ the radar carrier frequency. This equation can be used~\cite{davies} to obtain the index of refraction as a function of the plasma frequency:
\begin{equation}
\label{eq:index}
n^2 = 1-\frac{N e^2}{\freqradm^2 \epsilon_{0}m} = 1-\frac{\pfreqradm^2}{\freqradm^2}.
\end{equation}
In this low density plasma, collisions between free electrons and molecules, as well as geomagnetic effects are neglected. The plasma frequency is
\begin{equation}
\label{eqn:plasma}
\pfreqradm \equiv \sqrt{\frac{N e^2}{\epsilon_0 m}} 
\end{equation}

The index of refraction can be either real or imaginary. If $\pfreqradm > \freqradm$ (overdense), $n$ is imaginary and the waves cannot penetrate the medium. Specular reflection is the typical scattering regime for overdense plasma bodies. In the case $\pfreqradm < \freqradm$ (underdense), waves penetrate the medium. Scattering in the underdense regime is primarily via Thomson scattering~\cite{jackson} which, depending on the size of the scattering body relative to the wavelength, can interfere either constructively or deconstructively at the receiver. 

\begin{figure}[h]
\centerline{\includegraphics[trim=0.0in 0.1in 0.5in 0.4in, clip=true, width=0.45\textwidth]{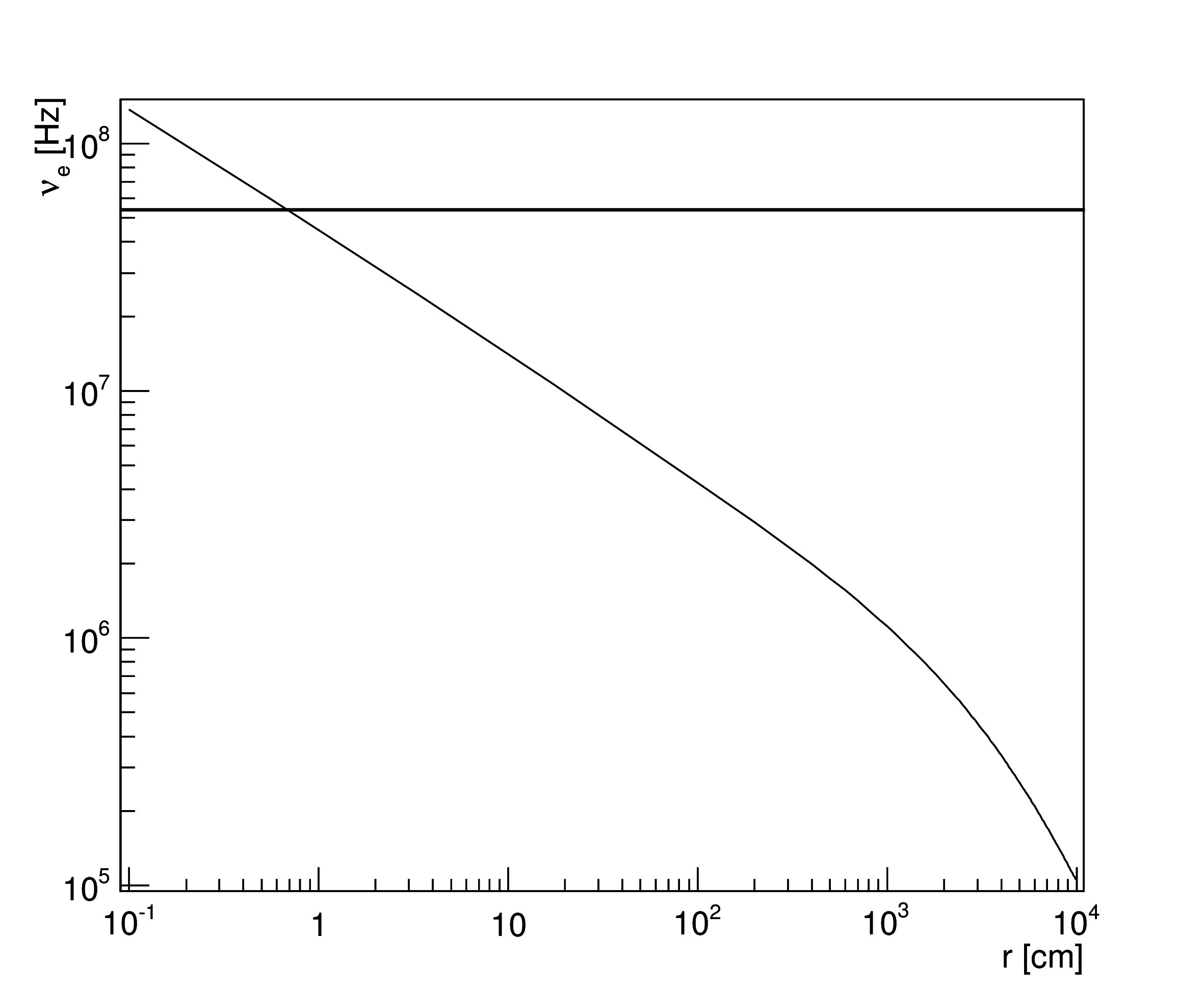}} 
{\caption{Plasma frequency as a function of radius at \xmax\ for a $10^{19}$~eV shower calculated using Gaisser-Hillas and NKG parameterizations. Gaisser-Hillas parameters are averages of values obtained by CORSIKA simulations. The horizontal black line corresponds to the TARA radar carrier frequency at 54.1~MHz.}
\label{fig:plasmafreq}}
\end{figure}

Figure~\ref{fig:plasmafreq} shows the plasma frequency as a function of radius, calculated from Gaisser-Hillas and NKG in a manner similar to the values obtained in Figure~\ref{fig:odcomp}. Given the TARA 54.1~MHz radar carrier, the shower only appears overdense at small radii, $\sim 1$~cm. 

\subsection{Collisional Damping}
\label{sub:collis_damp}

Collisions between free electrons and atmospheric molecules will tend to damp reradiation as the electrons' motion becomes incoherent. Scattered power can be expected to decrease as the effective collision frequency $\nu$ increases, particularly when $\nu$ becomes large compared to the sounding frequency. We can modify Equation~\ref{eq:pfreq_eom} to include the effects of collisional damping by adding a term assuming that all of an electron's excess momentum gained from the radar carrier is lost upon collision with a molecule. 

\begin{equation}
\label{eq:eom}
e\mathbf{E} = m\frac{\partial^2 \mathbf{r}}{\partial t^2} - m \nu \frac{\partial \mathbf{r}}{\partial t}\,.
\end{equation}
The index of refraction is then given by 
\begin{equation}
\label{eq:nsq_coll}
n^2 = 1 - \frac{\pfreqradm^2}{\freqradm^2 (1-i\cfreqm/\freqradm)}\,.
\end{equation}

With collisional effects included, $n$ is thus complex. Specifically, $n \equiv \mu - i\chi$, with $\chi$ the absorption coefficient and $\mu$ the real part of the index of refraction. When collisions are included, there are no longer two distinct regimes; rather incident waves are both partially reflected and partially absorbed. 

\begin{figure}[h]
\centerline{\includegraphics[trim=0.0in 0.0in 0.0in 0.0in, clip=true, width=0.5\textwidth]{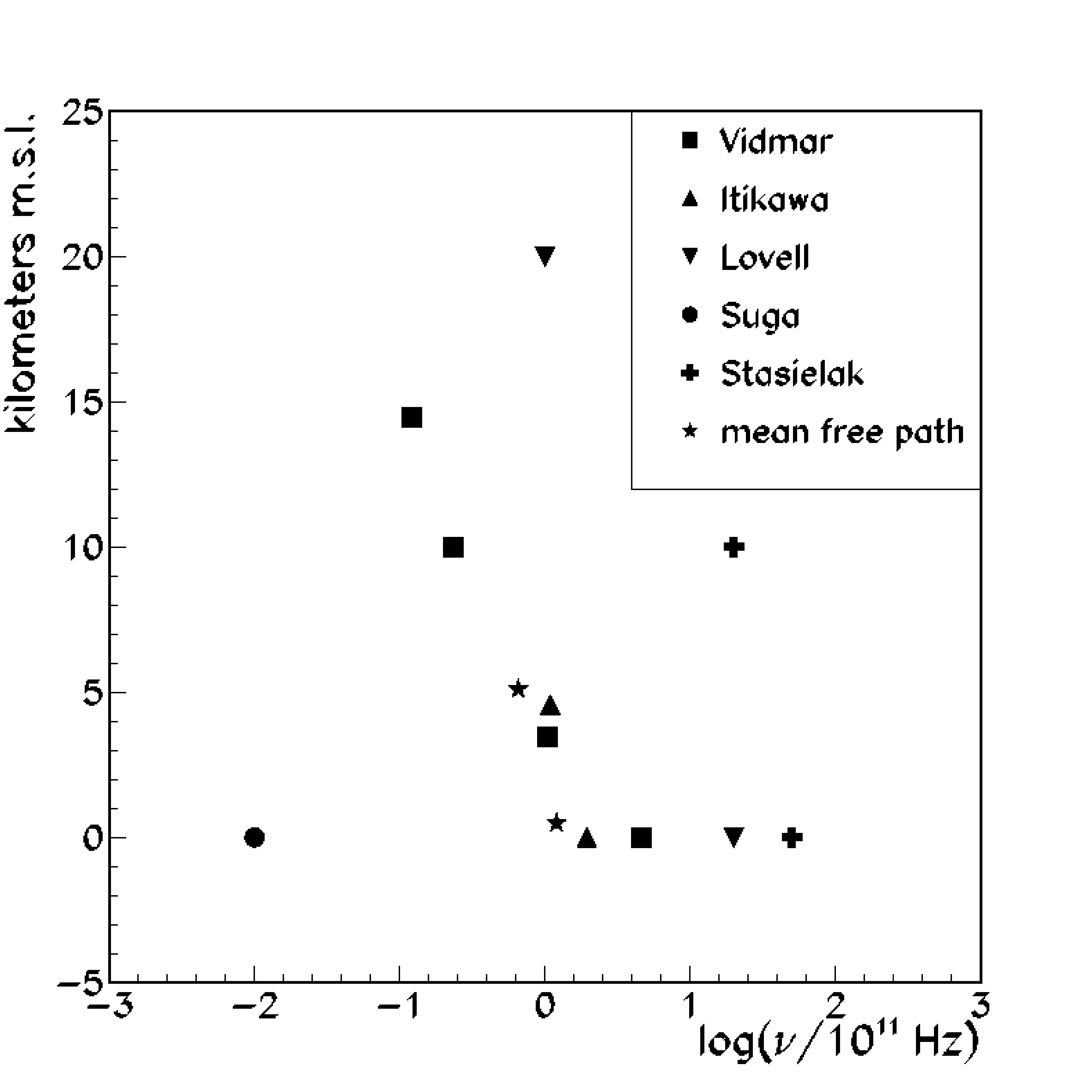}} 
{\caption{Survey of estimates of electron-neutral collision frequency as a function of altitude, mean sea level. Data points are from Vidmar~\cite{vidmar}, Itikawa~\cite{itikawa1971}, Lovell~\cite{belcorr}, Suga~\cite{suga}, and Stasielak {\em et al.}~\cite{stasielak2015}. A simple estimate is also shown, the result of dividing mean electron speed by the mean free path.}
\label{fig:nu}}
\end{figure}

Figure~\ref{fig:nu} gives values of $\nu$ for atmospheric plasmas as calculated by several authors, as well as a simple estimate obtained by dividing the mean thermalized electron velocity by the mean free path. 

Not including the simple estimate, all values were calculated as functions of the momentum transfer cross-section and electron velocity~\cite{shkarofsky1961_cfreq}: 
\begin{equation}
\label{eq:cfreq_ptrans}
\cfreqm(\text{v}) = \text{v}\left(N_{\text{N}_2}Q_m^{\text{N}_2}(\text{v}) + N_{\text{O}_2}Q_m^{\text{O}_2}(\text{v})\right)\,.
\end{equation}
$Q_m(\text{v})$ is the momentum transfer cross-section as a function of velocity $\text{v}$ and $N$ is the number density. 

By definition, the average shower particle energy at \xmax\ in air is $E_{\text{c}} = 81$~MeV. 
Electron temperatures up to 1000~K ($\langle E \rangle = \frac{3}{2}\langle k_{B}T \rangle$) were used in Equation~\ref{eq:cfreq_ptrans} to calculate the resultant values, which corresponds to $\sim 0.1$~eV. Ionization electron energies increase with that of incident particles~\cite{itikawa_n2cross} (shower electrons), therefore the electron velocities used to calculate the values in Figure~\ref{fig:nu} are likely too low. 

Above 1~keV, $Q_m^{\text{N}_2}$ decreases nearly two orders of magnitude for each decade increase in electron energy~\cite{itikawa_n2cross}. Data for electron energies in the range of interest ($\sim 10$~MeV) do not exist in the literature. If the trend of decreasing momentum transfer cross-section continues, then effective collision frequency values in Figure~\ref{fig:nu} are overestimates, caused by assuming a near-thermal secondary electron energy distribution. $N_{\text{O}_2}$ is small compared to $N_{\text{N}_2}$ and the momentum transfer cross-section of oxygen molecules appears to flatten as it approaches 1~keV~\cite{itikawa_o2cross}. Therefore, it is unlikely the oxygen term in Equation~\ref{eq:cfreq_ptrans} will compete with the nitrogen term.

Without the benefit of high-energy electron data, one can estimate the effect of collisional damping assuming that the collision rate in air is $\sim 10^{11}$~Hz. Comparing Equations~\ref{eq:index}~and~\ref{eq:nsq_coll}, an effective mass can be defined as $m_{\text{eff}} = m\left(1 - i\frac{\cfreqm}{\freqradm}\right)$. Scattering power is proportional to $(\partial^2 \mathbf{r}/\partial t^{2})^{2} \propto 1/m^{2}$. The damping factor is 
\begin{equation}
\label{eq:damping_mag}
\left|\frac{m_{\text{eff}}}{m}\right|^2 \simeq \left(\frac{\cfreqm}{\freqradm}\right)^2\,.
\end{equation}
Given $\freqradm = 10^8$~Hz and a possibly overestimated collision frequency $\cfreqm = 10^{11}$~Hz, the damping factor is $10^{6}$. The damping phenomenon is potentially catastrophic to the radar detection scheme, a fact that has been neglected in recent radar detection literature~\cite{gorham}.

\begin{figure}[t]
\centerline{\includegraphics[trim=0.1in 0.2in 0.5in 0.0in, clip=true, width=0.5\textwidth]{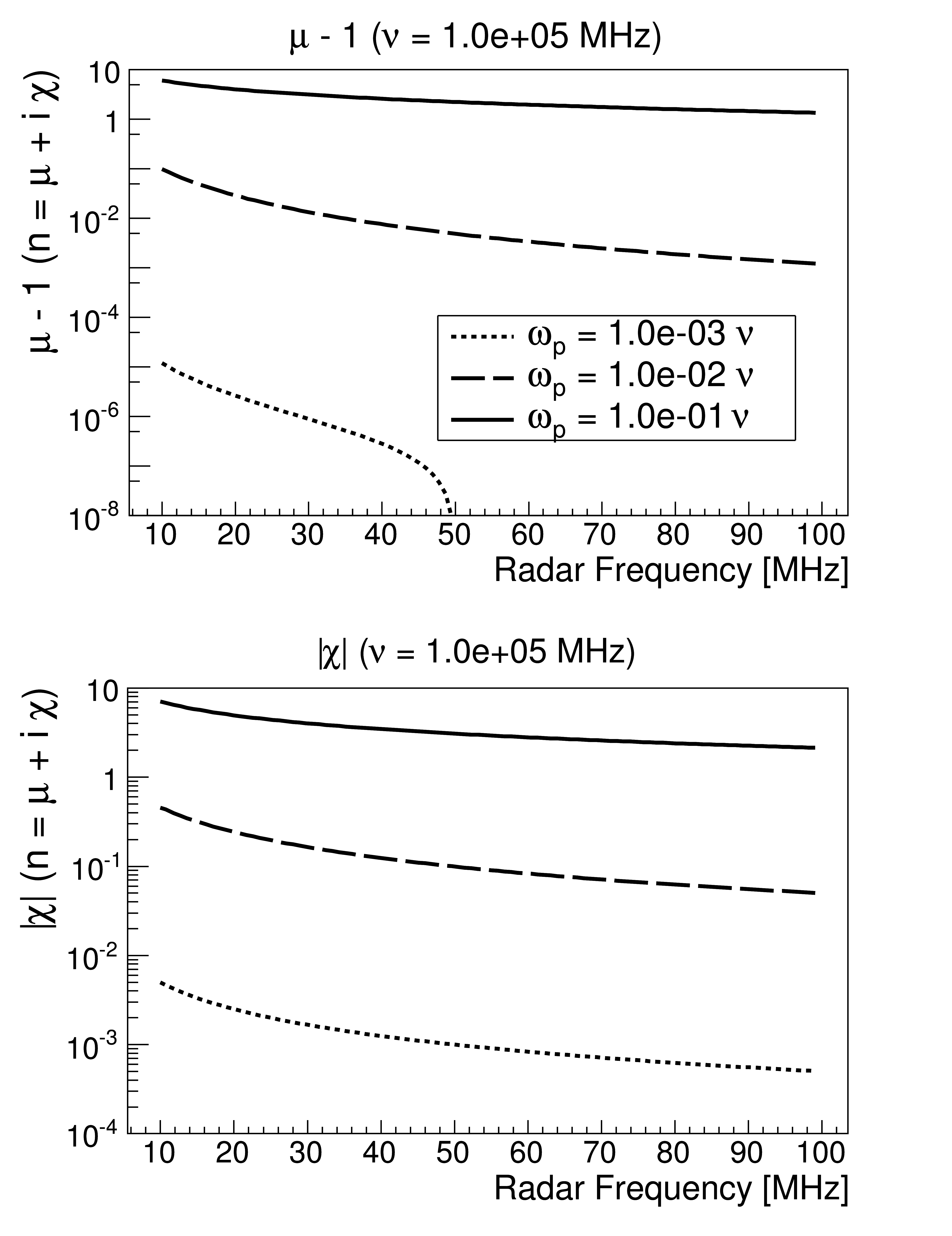}}
{\caption{Real and imaginary parts of the index of refraction ($n=\mu-i\,\chi$) with $10^{11}$~Hz collision frequency each with $\pfreqradm = (10^{-3},\;10^{-2},\;10^{-1})\,\cfreqm$. $\mu - 1$ is plotted to emphasize the difference between the $10^{-2}\,\nu$ and $10^{-3}\,\nu$ curves which are very close to unity. The TARA radar frequency is 54.1~MHz.}
\label{fig:ior}}
\end{figure}

Figure~\ref{fig:ior} shows $\mu$ and $\chi$ as a function of radar carrier frequency near 54.1~MHz at three different plasma frequencies $\pfreqradm = (10^{-3},\;10^{-2},\;10^{-1})\,\nu$, with $\nu = 10^{11}$~Hz. The refractive part of the index of refraction, $\mu$, does not deviate appreciably from unity at 54.1~MHz and plasma frequency \pfreqrad\ equal to $10^{-3}\,\cfreqm$. Absorption in the medium is of order $10^{-3}$. Thus we might expect that scattering power is very small even when the plasma frequency meets and exceeds the radar frequency ($\pfreqradm \simeq 10^{-3}\;\nu$). 

\subsection{Forward-scattering Enhancement}
\label{sub:forward}
\begin{figure}[t]
\centerline{\includegraphics[trim=0.0in 0.2in 0.68in 0.3in, clip=true, width=0.5\textwidth]{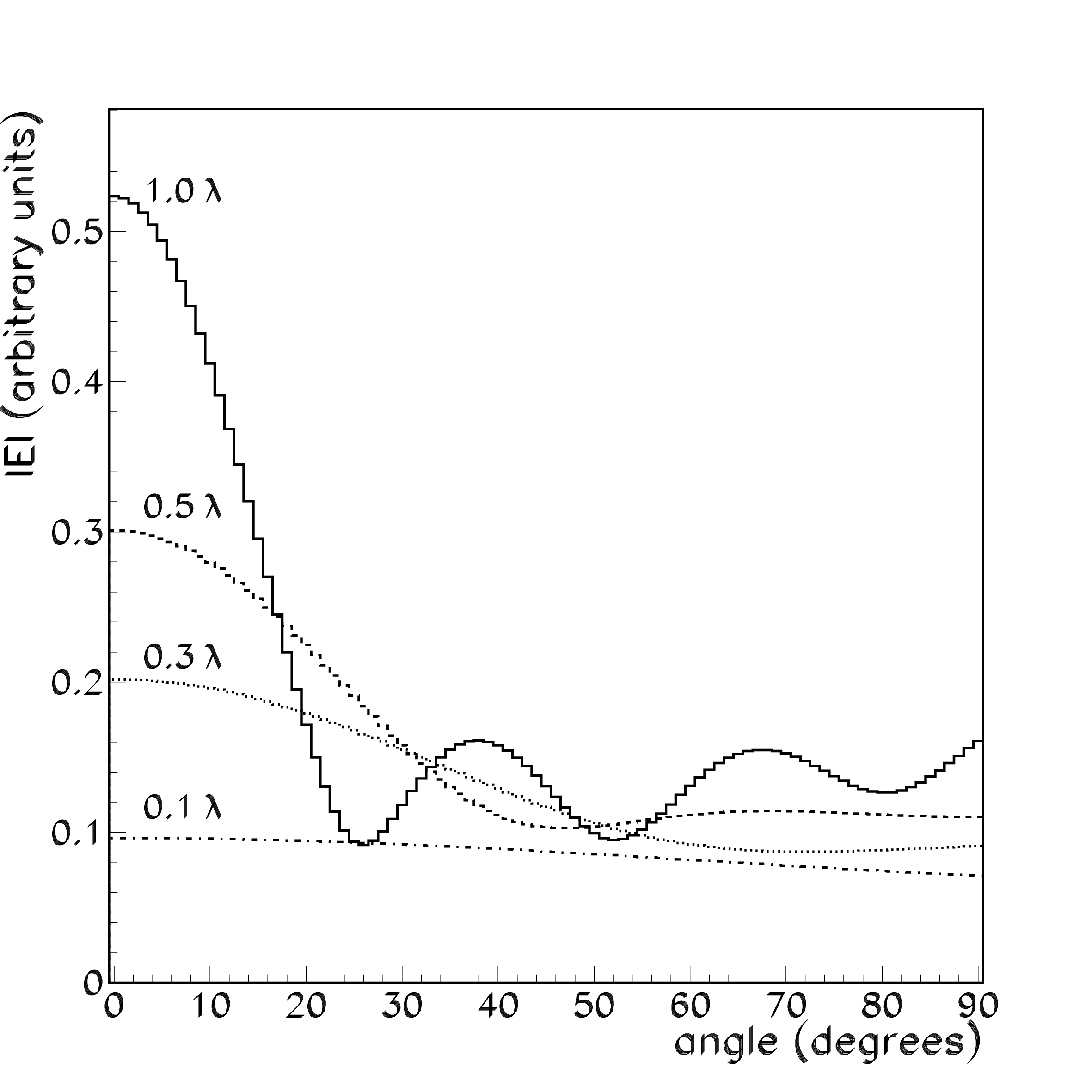}} 
{\caption{Relative scattered electric field~\cite{king} magnitude for several different cylinder radii, expressed as fractions of the incident wavelength, as a function of angular deviation from the forward scattering direction.}
\label{fig:cylinder}}
\end{figure}

The relative size of the radar wavelength $\lambda$ to the characteristic lateral size of the target $a$ determines if there is a peak in the scattered radiation pattern. Figure~\ref{fig:cylinder} shows a comparison of metallic cylinder forward-scattering radiation patterns as a function of the angle from the forward direction for several different cylinder radii $a$ expressed as fractions of the radar wavelength. The diffraction peak vanishes when $a/\lambda < 0.1$. 

If the majority of scattered power comes from near the shower core in the region that would be considered overdense in the collisionless regime, the scattering region with radius $a\sim 1\,\text{cm}$ can be represented as a thin conductive wire with no forward enhancement. In the opposite case, one considers a cylinder with radius equal to $\lambda$, which encloses over 30\% of shower electrons. This scattering volume is expected to have appreciable forward enhancement. 

\subsection{Simulation}
\label{sub:sim}
Energetic shower particles travel in a disk coaxial with the EAS direction of propagation.  Less energetic products are quickly thermalized on a time scale of the order of the free electron lifetime.  From the perspective of a radar system, an EAS is a (narrow) disk moving through the atmosphere at the speed of light, leaving a quickly fading plasma trail. The simulation of radar echoes is made possible using the bi-static radar equation, TARA detector geometry and known EAS dynamics to calculate the received signal based on a scattering model, as detailed below. 

\begin{figure}[t]
\centerline{\includegraphics[trim=0.5in 0.4in 0.4in 0.5in, clip=true, width=0.5\textwidth]{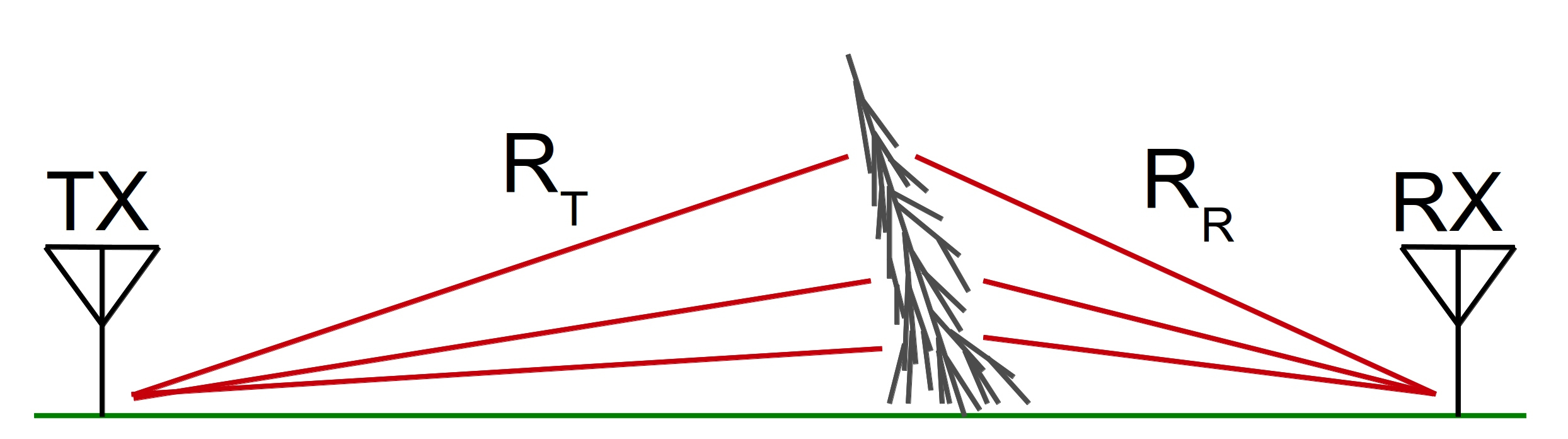}}
{\caption{Contributions from paths of varying lengths, $\text{TX}\rightarrow \text{target} \rightarrow \text{RX}$, summed at the receiver result in a chirp signal (bi-static configuration).}
\label{fig:superposition}}
\end{figure}

The superposition principle implies that multiple scattering path lengths from different segments of the shower track (see Figure~\ref{fig:superposition}) result in summation of scattered rays of the same frequency but with different phase. Unlike other radar applications, the target is moving at essentially the same speed as the interrogating wave, so phase evolves rapidly. EAS radar echoes exhibit rapidly changing frequency~\cite{underwood}. Such signals are often referred to as ``chirps''.

During a given time step \dt\ at time $t_{i}$, the bi-static radar equation is applied to each longitudinal shower segment (each with length $c\, \dtm$) from which light could have reached the receiver. A segment $j$ is included if the shower progress $\mathcal{P}_{i}$, the distance the shower has travelled since first interaction, is greater than the distance from segment $j$ to the receiver $R_{R,j}$ minus the progress of the shower at segment $j$, $\mathcal{P}_{j}$. The plasma state of segment $j$, and thus RCS properties, are that of the segment at retarded time $t'_{i,j} = (\mathcal{P}_{i} - R_{R,j} - \mathcal{P}_{j})/c$. The time $(\mathcal{P}_{i} - R_{R,j})/c$ is defined by scattered light from segment $j$ leaving the segment. It does not include shower age which must be tracked carefully to include the effect of plasma dissipation.

Each segment has magnitude and phase determined by the segment's total path length, radar wavelength, RCS at the current point in the shower, shower geometry relative to the TX and RX and geometrical antenna factors.  The integrated contribution of all segments $j$ is the received signal at time $t_{i}$
\begin{equation}
\label{eq:simsum}
V_{R,i} = \sum_{j} \sqrt{P_{j}(t_{i}')\, Z}\;.
\end{equation}
$P_{j}(t_{i}')$ is the received power from the bi-static radar equation (Equation~\ref{eq:bi}) calculated at the retarded time, and $Z$ is the impedance of the receiving antenna. The sum should properly account for phase factors in $P_{j}$. In practice, antenna impedance $Z$ is a function of frequency. Here, a fixed value is assumed. Error introduced from this assumption is discussed in Section~\ref{sub:uncertainty}.  

Signal components scattered early in shower evolution when the RCS is small and path length is large will have low amplitude, while segments near \xmax\ (typically 3-5~km from ground level at the location of the detector) have larger RCS due to high plasma density near \xmax\ and have shorter path lengths so the expected received signal later in the shower is greater.

Several physical models are included in the simulation to increase accuracy. Gaisser-Hillas parameters are coded as functions of primary energy \enot\ using CORSIKA Monte Carlo data~\cite{zech}. The NKG function combined with Nerling's parameterization~\cite{nerling} of $\alpha$ describe the ionization lateral distribution. Atmospheric density as a function of altitude is obtained from the 1976 Standard Atmosphere~\cite{atmosphere}. Electron lifetime as a function of altitude comes from Vidmar~\cite{vidmar}. Transmitter and receiver radiation patterns have been simulated~\cite{nec} and confirmed observationally (see~\cite{taranim}).

The electron lifetime my be underestimated due to the assumption of near-thermal electrons. This is the analog to overestimation of effective collision frequency in the Section~\ref{sub:plasma_scatt}. Electron lifetime for near-thermal electrons (500~K) is $\tau \simeq 10$~ns~\cite{vidmar,dogariu2013}. Electrons with greater energy require additional collisions with neutral molecules before attachment/recombination can occur. Without strong evidence in favor of decreasing momentum-transfer cross section, near-thermal electron lifetime is used in the simulation. Underestimated electron lifetime results in lower predictions for received power. 

\subsubsection{The Thin Wire Model}
\label{subsub:thinwire}

In order to examine the radar target presented by particular air showers, we adopt a simplified ``thin wire model'' in terms of which we can quantify the RCS. 

As described in Section~\ref{sub:forward}, the bulk of coherent scattered power will occur in a cylindrical region with radius less than $\lambda$, possibly much less due to the steep shower density profile. Only the narrow overdense region (collisionless model) is used in the simulation. It may be regarded as a short-lived, thin conductive wire. The overdense region radius is determined from the lateral distribution, which is used in the thin-wire approximation to give the segment cross-section. Forward scattering enhancement (Section~\ref{sub:forward}) is not expected from a scattering region with radius much less than $\lambda$. Neglecting the possibility of such enhancement conservatively underestimates received power, thus preventing underestimation of the RCS upper limit discussed below in Section~\ref{sec:rcscalc}.

The RCS of a perfectly conducting thin wire $\sigma_{\text{TW}}$ several wavelengths long but only a fraction of a wavelength in diameter is given by~\cite{crispin},
\begin{equation}
\label{eq:thinwire}
\sigma_{\text{TW}} = \frac{\pi L^2 \, \text{sin}^2 \,\theta \left[\frac{\text{sin}\,\eta}{\eta}\right]^2}{\left(\frac{\pi}{2}\right)^2 + \left(\text{ln}\frac{\lambda}{\gamma \pi a \text{sin}\,\theta} \right)^2}\, \text{cos}^4\,\phi\;,\,\eta = \frac{2 \pi L}{\lambda}\,\text{cos}\,\theta\,.
\end{equation}
$L$ is the wire length, $\theta$ is the angle between the wire and the direction of incidence, $\phi$ is the angle between the incident wave polarization and the wire axis, $a$ is the radius, $\lambda$ is the wavelength and $\gamma$ is 1.78, $e$ raised to the power of Euler's constant 0.577. 

In the simulation, the thin-wire radius $a$ is dependent on the shower lateral distribution, and the state of the plasma at retarded time $t'$. Segment $j$ at time step $i$ is described by NKG with $N = N_{0,j}\,\text{exp}(-t_{i,j}'/\tau)$, where $N_{0,j}$ is the initial number of ionization electrons in segment $j$ and $\tau$ is electron lifetime. The thin-wire radius $a$ is the largest radius for a given shower segment within which the free electron density exceeds the density that corresponds to a plasma frequency of 54.1~MHz. 

We note the following features of the thin-wire model~\cite{crispin} and Equation~\ref{eq:thinwire}:

\begin{itemize}

\item The dependence on polarization enters as $\cos^{4}{\phi}$, peaked for polarization parallel to the wire axis and zero for polarization perpendicular to wire axis. To maximize received signal, {\bf E}-field polarization should be parallel to the air shower trajectory. 

\item The radius $a$ of the wire contributes logarithmically to the radar cross section. Since the radius of the overdense region is linearly dependent on \enot, the RCS will be only logarithmically dependent on primary energy and not proportional to primary energy as might be expected in a simpler model. 

\item Unlike scattering in the $a \approx \lambda$ regime, scattered radiation will not be enhanced in the forward direction (Figure~\ref{fig:cylinder}). Rather, short-thin-wire radiation is treated as dipole emission.  

\item A change of $\sim 30^{\circ}$ in aspect $\theta$ will cause significant oscillations in $\sigma_{\text{TW}}$ because of amplification in the $(\text{sin}\,\eta/\eta)^2$ term.

\end{itemize}

Known shortcomings of the model include:

\begin{itemize}

\item High electron density air shower core ionization will likely be an imperfect conductor, due again to the high rate of collisions with neutral molecules. Significant damping of RCS is expected due to this effect. 

\item Even absent collisional effects, plasmas have an associated ``skin depth" given by $\delta = c/\omega_p$. If the radius of the wire $a \ll \delta$, only a small part of the incident radiation may be absorbed and re-radiated by the wire.

\item Contributions from parts of the shower with radii greater than $a$ are neglected.

\end{itemize}

\subsubsection{Frequency Shift: The ``Chirp''}
\label{subsub:chirp}

Some of the characteristics of the chirp can be understood using geometrical arguments. Let $R$ be the total path length $R_{T}+R_{R}$ and $\dot{R}=dR/dt$ the instantaneous rate of change of path length. $\dot{R}$ is proportional to chirp rate.  If $\dot{R} > 0$ the received frequency is less than the radar frequency and if $\dot{R} < 0$ the received frequency is greater than the radar frequency. Time-dependent frequency is analogous to Doppler shifted light or sound waves, though not identical because the wavelength of scattered radar carrier is fixed at the receiver---only combined phase changes.

CR air showers which start far away and move toward the Earth's surface produce ``down chirps'' (decreasing frequency) starting above the carrier frequency as long as the core location is between TX and RX.  There are geometries in which $\dot{R}$ approaches zero, then becomes positive as the shower crosses the line connecting TX and RX.  In this case, the chirp frequency will first descend to, then below the radar frequency. Neutrino air showers originating close to the earth's surface or low in the atmosphere could produce down chirps that start below the radar frequency and descend to lower frequencies.

The chirp signature contains information about air shower geometry. With the exception of lateral symmetry about a plane perpendicular to the ground and containing the TX and RX points, and a rotational symmetry about a line connecting the transmitter and receiver, radar echoes are unique.

\begin{figure}[!h]
\centerline{\includegraphics[trim=0.0in 0.23in 1.2in 0.3in, clip=true, width=0.5\textwidth]{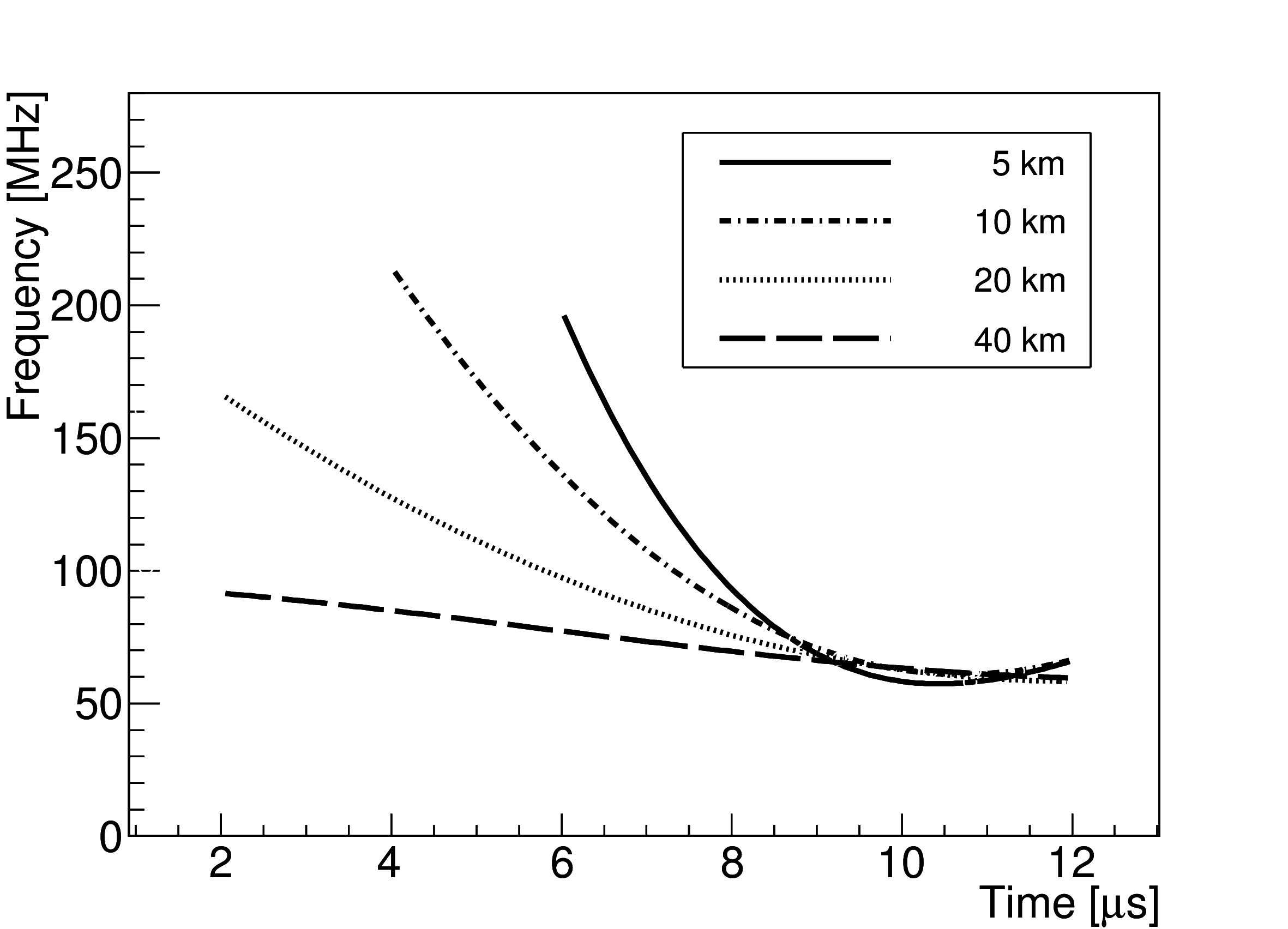}} 
{\caption{Simulated chirp spectra fits to highest amplitude frequency component for four different geometries. Each simulation represents a vertical, 10~EeV CR shower located midway between transmitter and receiver. $\text{TX}\rightarrow \text{RX}$ separation distances are shown on the legend. Both the time offsets and absolute frequency ranges have been justified for direct chirp rate comparison.}
\label{fig:txrxsep}}
\end{figure}

Figure~\ref{fig:txrxsep} shows frequency vs. time for four canonical chirps with different $\text{TX}/\text{RX}$ separation distances.  Canonical chirps are simulated radar echoes from small zenith angle, 10~EeV air showers located midway between the transmitter and the receiver. The curves represent the highest amplitude frequency component in each time bin. In the following sections, duration is defined as the total time during which received power is within 10~dB of maximum. Bandwidth is the frequency difference between the two maximum$-10$~dB power points, on either side of the frequency at maximum power.  The plot indicates that echoes with small $\text{TX}/\text{RX}$ separation distances are short in duration and have large chirp rates. 

\begin{figure}[!h]
\centerline{\includegraphics[trim=0.0in 0.25in 0.0in 0.3in, clip=true, width=0.4\textwidth]{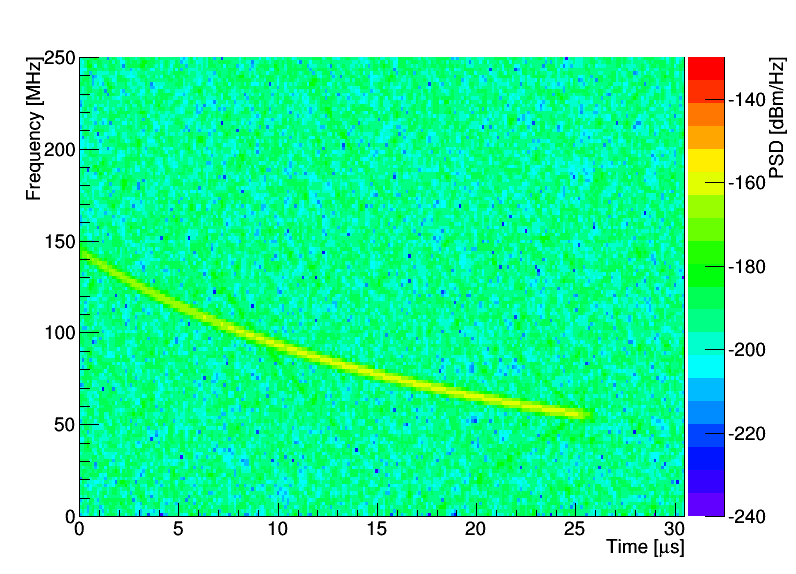}} 
\centerline{\includegraphics[trim=0.0in 0.25in 0.0in 0.3in, clip=true, width=0.4\textwidth]{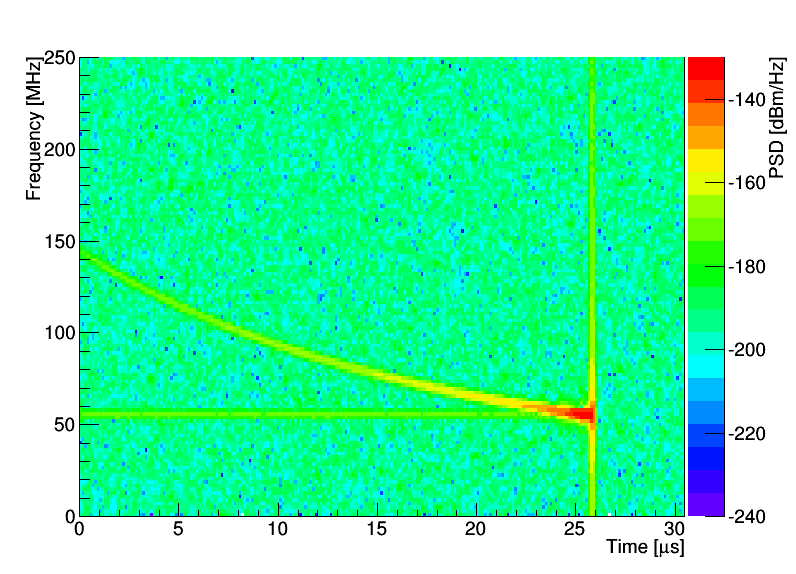}} 
\centerline{\includegraphics[trim=0.0in 0.25in 0.0in 0.3in, clip=true, width=0.4\textwidth]{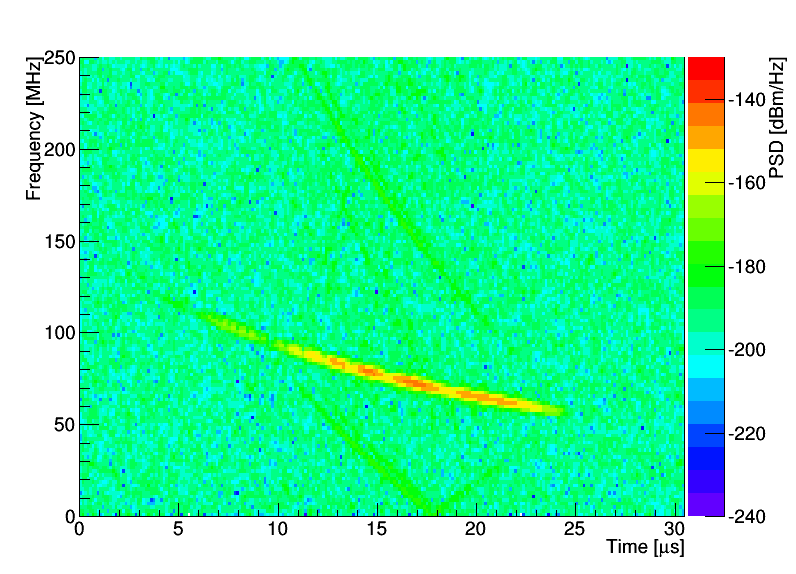}} 
{\caption{Spectrograms showing simulated radar echoes for a shower midway between transmitter and receiver and inclined 30$^{\circ}$ out of the TX/RX plane. Top: Electron lifetime is fixed at 1~ns and antenna gain is held constant. This configuration simulates a small scattering object travelling at the speed of light toward the ground. Middle: Electron lifetime is fixed at 100,000~ns and antenna gain is held constant. This configuration simulates a scattering rod beginning high in the atmosphere and growing toward the ground at the speed of light. Bottom: Electron lifetime is determined from empirical models and RCS comes from the thin-wire approximation and shower evolution models. Antenna gain is determined from lookup tables generated by NEC~\cite{nec}. This configuration simulates a cosmic ray radar echo.}
\label{fig:const_fvst}}
\end{figure}

A simple simulation that only tracks the phase of a point source with speed $c$ will correctly yield the chirp signature of a given geometry. This fact is important because the assumption of a scattering model does not affect the frequency vs. time graph. Three radar echoes have been simulated with the same geometry and different electron lifetimes. Spectrograms of the simulated waveforms are shown in Figure~\ref{fig:const_fvst}. The first two have constant cross-sections; each segment has the same value of $\sigma$ in the bi-static radar equation. The third uses the thin-wire approximation described above for the RCS, the parameters of which are informed by air shower evolution models and evolving electron lifetime~\cite{vidmar}. 

\begin{figure}[h]
\centerline{\includegraphics[trim=0.0in 0.0in 0.0in 0.2in, clip=true, width=0.5\textwidth]{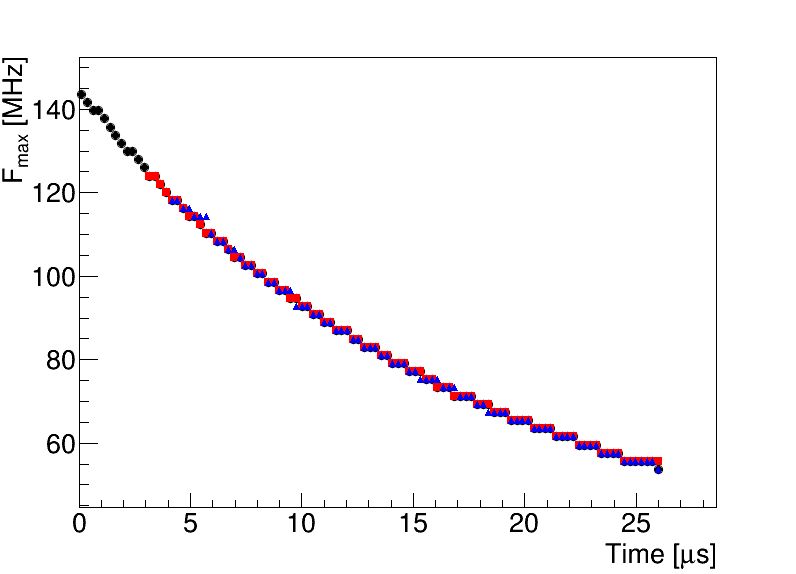}} 
{\caption{Plot showing \fmax\ vs. time for the three simulated echo waveforms shown in Figure~\ref{fig:const_fvst}. Black points represent \fmax\ for the short lifetime waveform. Red and blue points represent \fmax\ for the long lifetime and full simulation waveforms, respectively.}
\label{fig:const_fvst_comp}}
\end{figure}

Simulation 1 (top) has a very short, fixed, free electron lifetime, similar to a very small metallic sphere moving with speed $c$. There is only one scattering path length per time step. Simulation 2 (middle) has a lifetime longer than the travel time of the shower from first interaction to ground level, similar to a thin metallic cylinder that grows toward the ground at speed $c$. Simulation 3 is the output of the full simulation including lookup tables for the TARA TX and RX antennas in three dimensions, shower evolution models (Gaisser-Hillas, NKG, Nerling, Standard Atmosphere, etc., as above), and the thin-wire approximation for the RCS. The frequency vs. time graphs for the three simulations are superimposed in Figure~\ref{fig:const_fvst_comp}, where only the maximum frequency in a given time bin is plotted. This comparison shows that frequency vs. time is independent of the radar target.

\begin{figure}[!h]
\centerline{\includegraphics[trim=0.0in 0.0in 0.0in 0.2in, clip=true, width=0.5\textwidth]{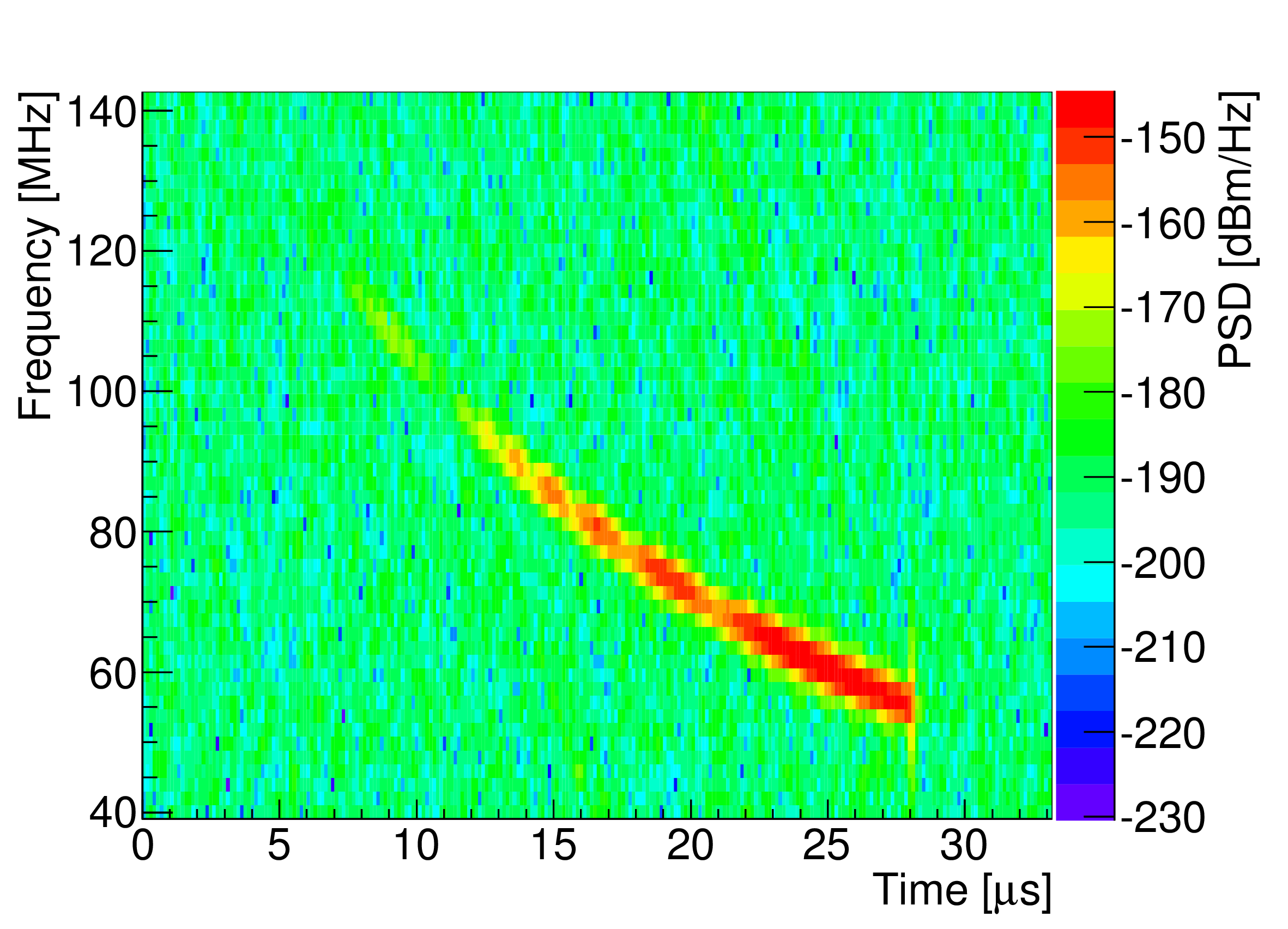}} 
{\caption{Radar echo spectrogram from a $10^{19}$~eV shower located midway between transmitter and receiver inclined $30^{\circ}$ out of the plane connecting the two.}
\label{fig:chirp}}
\end{figure}

Figure~\ref{fig:chirp} shows the spectrogram of a canonical shower generated with the full TARA detector simulation. Properties relevant to detection are chirp slope ($\sim 2$~MHz/$\mu$s), frequency at maximum power \fmax (near 60~MHz), and chirp duration ($\sim 10$~$\mu$s).

\section{Radar Receiver and DAQ}
\label{sec:data}

The TARA DAQ (described in detail in~\cite{taranim}) records custom binary files that each contain a metadata header and 1,000 triggers. DAQ settings are contained in the metadata header and each trigger includes its own header and four waveforms, each 130.976~$\mu$s in duration, or 32,744 samples at 250~MS/s, composed of 16 bit signed integer samples. Each waveform corresponds to one of four DAQ input channels, which are fed by two dual-polarized log periodic dipole antennas (LPDA). While collecting data used in the present analysis, the TARA transmitter was operated with horizontal polarization and hence we are primarily concerned with the horizontal receiver channels in which the sensitivity will be greatest. 

Signals first pass through a series of RF filters and an amplifier on the front end of the RF DAQ system. Sequentially, the components are: lightning arrestor, which grounds the center conductor during periods of excessive voltage; RF limiter, which limits power to +11~dBm (decibels relative to mW); 30~dB amplifier; 40~MHz high pass (HP) filter, which attenuates frequency components below 40~MHz; FM band stop, which attenuates frequency components in the range 88--108~MHz; and a 90~MHz low pass (LP) filter, which attenuates frequencies above 90~MHz. 

A single binary output file contains both FD triggers and ``snapshot'' triggers. Snapshot triggers are taken automatically once per minute, and are used to estimate the noise background. The receiver is located on the campus of the Long Ridge fluorescence detector, located southwest of the TA surface detector. During FD data acquisition, trigger pulses are sent to the radio receiver DAQ. The hardware level trigger from the FD to the TARA DAQ forces one event and a time stamp to be recorded. The FD triggers used were low-level triggers, a subset of which corresponded to actual CR events. Star light intensity fluctuations, passing airplanes and self-calibration triggers are included in low-level hardware triggers. The rate of these triggers during standard FD acquisition is 3--5~Hz, much higher than expected for high energy CR events.

\begin{figure}[!h]
\centerline{\includegraphics[trim=0.0in 0.0in 0.0in 0.0in, clip=true, width=0.5\textwidth]{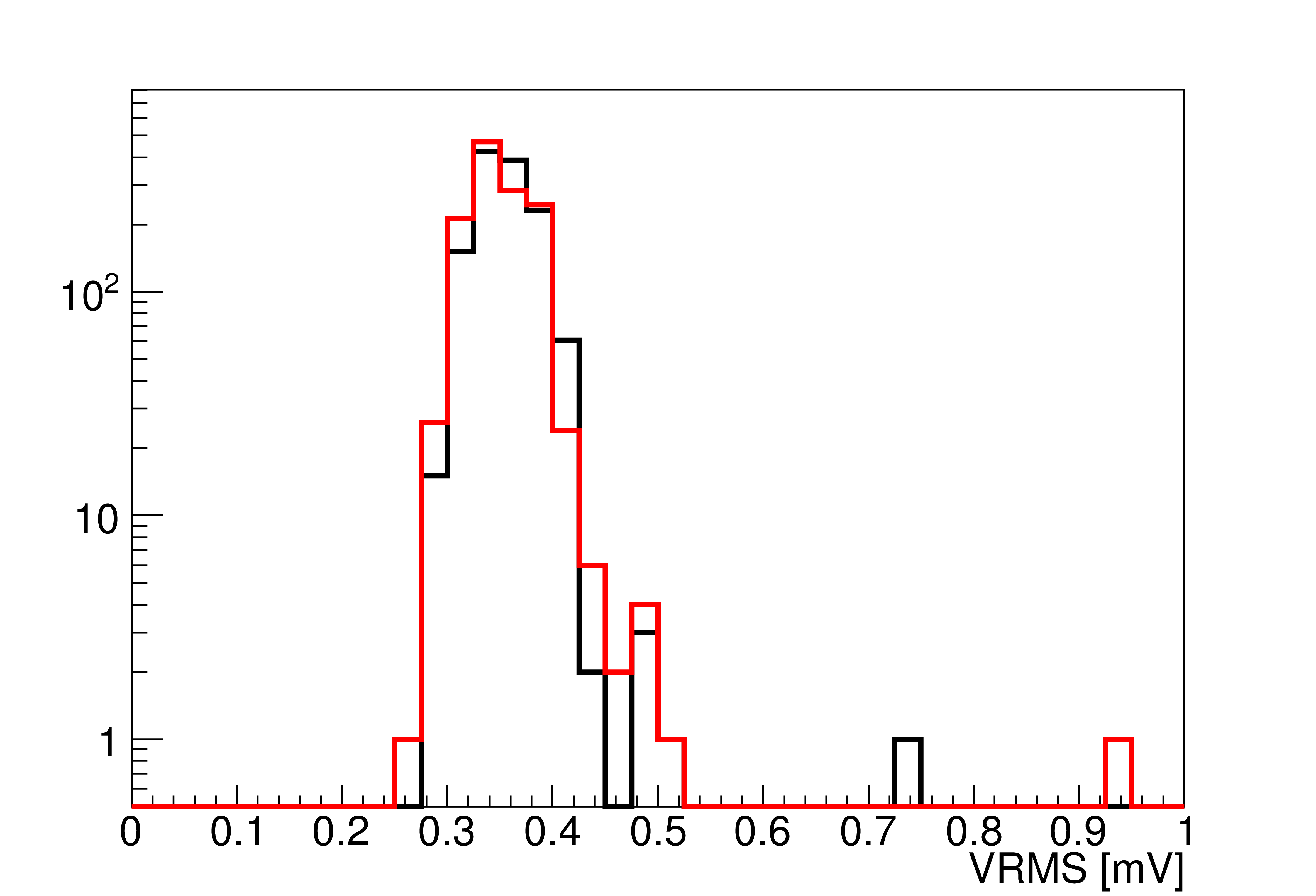}}
{\caption{Voltage RMS distribution of all matched FD-triggered events in black overlayed with that of snapshots in red, recorded during several FD observation periods from August, 2013 to April, 2014. Waveforms are notch- and HP-filtered before the RMS is calculated. A fixed number of matched events occur during each FD observation period. The number of snapshot RMS values included in the histogram, per run, is equal to the number of included FD trigger values.} 
\label{fig:fdtrigsnapcomp}}
\end{figure}

Snapshot triggers occur once per minute and are not correlated to any external trigger. These triggers offer an unbiased representation of the RF background, assuming there is no additional trigger noise from the FD electronics or TARA DAQ. A quantitative comparison of snapshot and FD triggers is shown in Figure~\ref{fig:fdtrigsnapcomp}. Recall that the majority of FD triggers are the result of low level FD trigger noise, and therefore should not have special waveform features. The black histogram is the distribution of VRMS values for all matched FD-trigger events. The red histogram is the distribution of snapshots selected for analysis (selection process described below) from the FD observation period. Good agreement between snapshot and FD-trigger RMS distributions indicates that there is not additional FD noise that can obscure signal. 

FD triggers and snapshots are used in the present analysis. FD triggers comprise a promising dataset because a subset of these triggers are radar data obtained during actual CR events in the field of view of the detector. It only operates when the moon is below the horizon on clear, dry nights. The average duty cycle is approximately 10\%. 

\begin{table*}[t]
\caption{Long Ridge FD-reconstructed event quality cuts. (NPE = number of photo-electrons; SDP = shower detector plane; see Figure~\ref{fig:fdgeo}. Pseudo-distance is the speed of light divided by the angular speed of the track.)}
\label{tab:fdcuts}
\begin{center}
\begin{tabular}{| c | c | p{8cm} |}
\hline
Cut Name & Criteria & Description \\ \hline \hline
good tube fraction & $\geq 3.5\%$ & data quality, noise reduction \\ \hline
good tubes & $\geq 6$ & data quality, noise reduction \\ \hline
NPE/degree & $\geq 25$ & data quality, sufficient signal \\ \hline
pseudo-distance & $\geq 1.5$~km & geometry, shower resolution \\ \hline
SDP angle & $< 80^{\circ}$ & geometry \\ \hline
$R_p$ & $\geq 1$~km & geometry \\ \hline
$\Psi$ & $0^{\circ} \leq \Psi < 150^{\circ}$ & geometry \\ \hline
$\delta\Psi$ & $< 36^{\circ}$ & fit reconstruction, $\Psi$ error \\ \hline
tangent fit & $\chi^2/\text{DOF} < 10$ & fit reconstruction \\ \hline
zenith angle & $< 70^{\circ}$ & geometry \\ \hline
$t_0$ & $< 25.6\,\mu$s & event occurs in trigger window \\ \hline
$R_p$ and $t_0$ & $(R_p > 5~\rm{km})$ or $(t_0 > 3~\mu{\rm s})$ & geometry \\ \hline
first tube depth & $150 < X_1 < 1200$~g/cm$^2$ & geometry, first tube illuminated from reasonable depth \\ \hline
observed slant depth & $>150$~g/cm$^2$ & geometry, minimum track length \\ \hline
\xmax\ &  $400 < \xmaxm < 1200$~g/cm$^2$ & fit reconstruction \\ \hline
\end{tabular}
\end{center}
\end{table*}

\section{Offline Processing}
\label{sec:offline}
\begin{figure}[h]
\centerline{\includegraphics[trim=0.0in 0.0in 0.0in 0.0in, clip=true, width=0.4\textwidth]{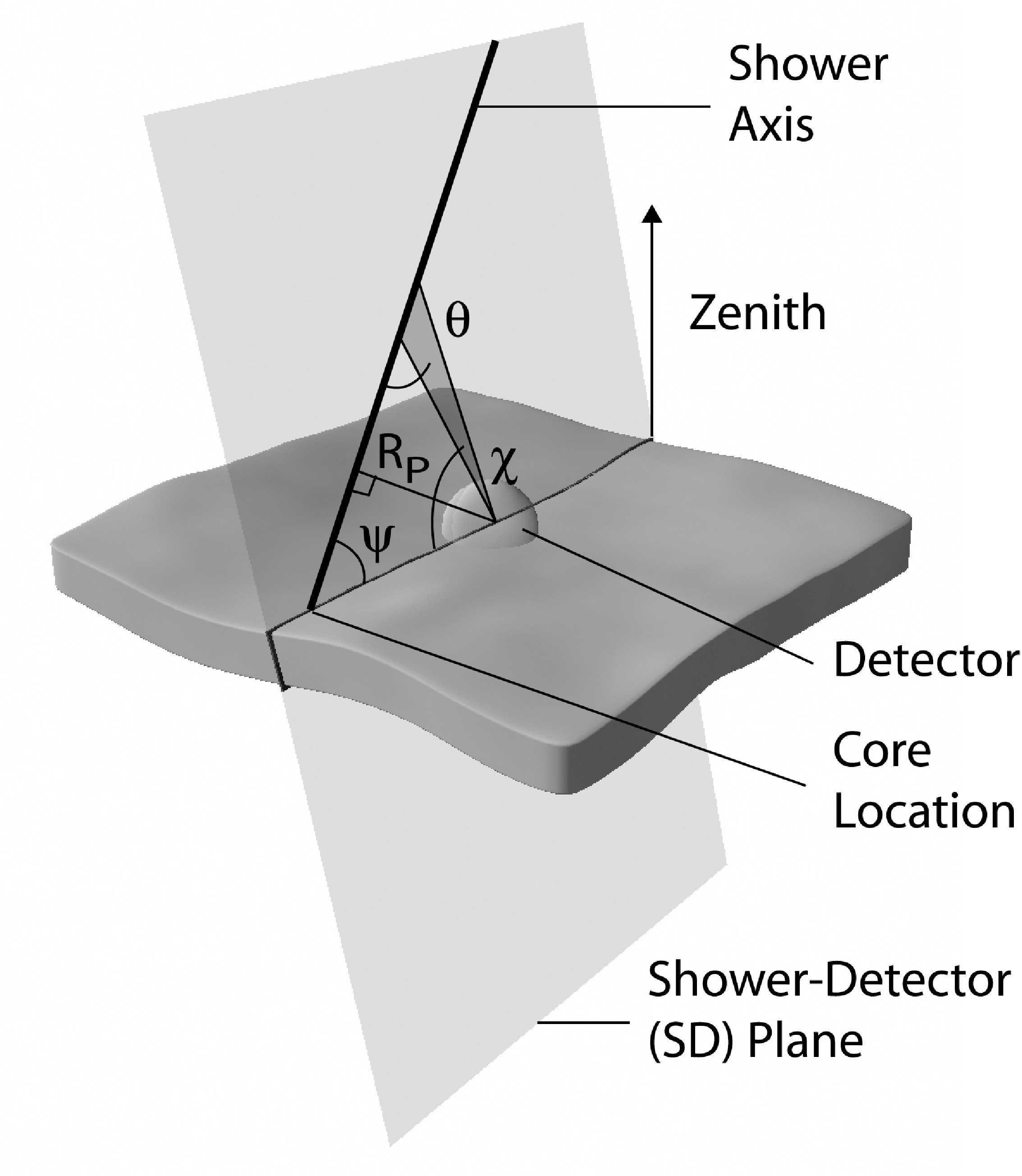}}
{\caption{Fluorescence detector geometry~\cite{stokes}.} 
\label{fig:fdgeo}}
\end{figure}

TARA events from FD triggers are time-matched with reconstructed Long Ridge FD events~\cite{abbasi2005FDmono} to remove noise triggers, after which the following reconstruction parameters are available: GPS time stamp, primary energy, core location, zenith/azimuth angle and \xmax. The reconstructed FD events from Long Ridge are selected by quality cuts designed to remove those which are reconstructed with large uncertainty. Table~\ref{tab:fdcuts} gives the list of cuts and their descriptions. Figure~\ref{fig:fdgeo} shows a diagram of the FD geometry.

\begin{figure}[h]
\centerline{\includegraphics[trim=0.0in 0.0in 0.0in 0.0in, clip=true, width=0.5\textwidth]{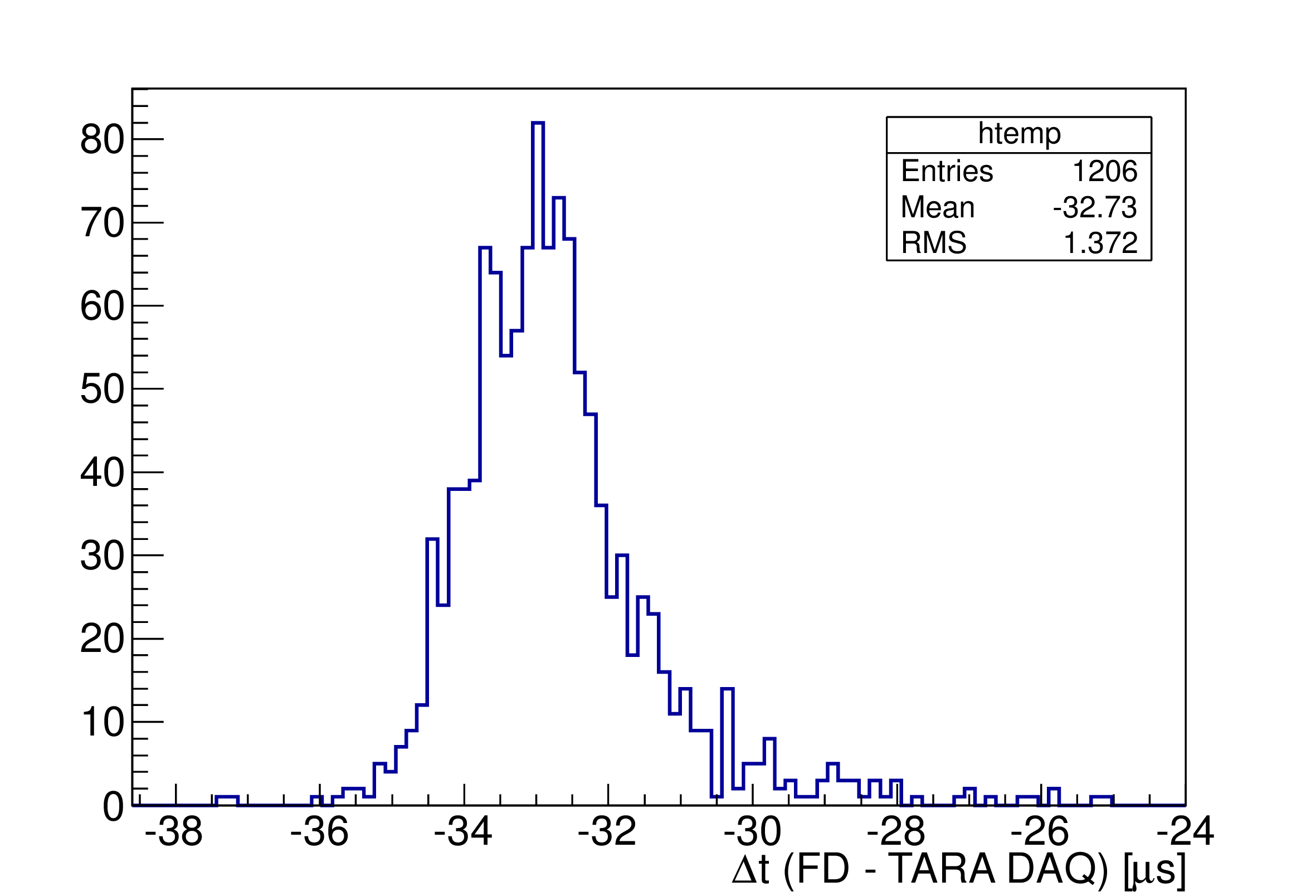}}
{\caption{Time difference between FD reconstructed events and the TARA FD-triggered events recorded during FD observation periods between August, 2013 and April, 2014. The 33~$\mu$s delay is caused by FD DAQ trigger formation, cable delay and TARA DAQ delay in signaling an event to the GPSY~\cite{gpsy2} GPS event logger.} 
\label{fig:dtall}}
\end{figure}

Triggers are matched with an absolute time difference less than 200~$\mu$s. Figure~\ref{fig:dtall} shows a histogram of the time difference between TARA FD triggers and reconstructed TA FD events collected over the course of several FD observation periods between August, 2013 and April, 2014. Matched FD-triggered events recorded during this date range are the focus of this analysis. Ignoring the delay in trigger formation, absolute timing uncertainty is 20~ns~\cite{gpsy2}.

\subsection{Trigger Time Range}
\label{sub:trig}
Spurious noise (discussed in Section~\ref{sub:filtering}) can interfere with the signal search and possibly mimic a positive detection. Therefore, a fixed time range of waveform samples is included in the analysis to reduce the probability that noise is present in the analyzed portion of the waveform. Any waveform samples or features outside this time range are not considered in any steps of the analysis. Only signals occurring between 48~$\mu$s before and 5~$\mu$s after the expected arrival time were considered. 

\subsection{Filtering}
\label{sub:filtering}
\begin{figure}[!h]
\centerline{\includegraphics[trim=0.0in 0.0in 0.0in 0.0in, clip=true, width=0.5\textwidth]{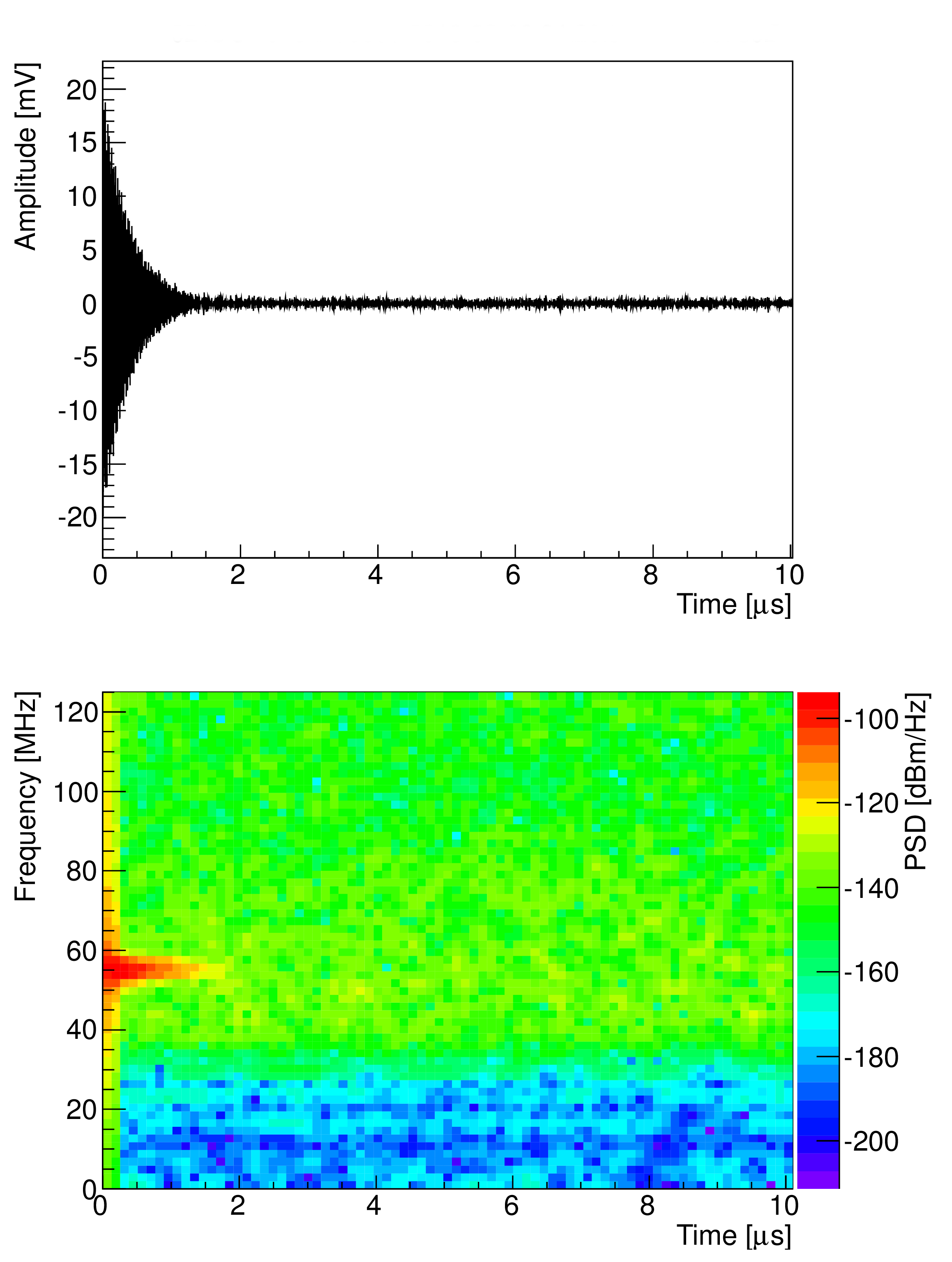}}
{\caption{Event display for a snapshot chosen at random from the August, 2013 FD run with both 54.1~MHz notch and 30~MHz high pass digital filters applied. Only the first 10~$\mu$s of the 132~$\mu$s waveform is shown to emphasize filter characteristics. The high pass filter is fixed and attenuates frequencies below approximately 30~MHz from the beginning of the waveform. Carrier amplitude decreases as the adaptive notch filter find the correct amplitude and phase.}
\label{fig:snapevdisp}}
\end{figure}

The radar carrier frequency component at 54.1~MHz dominates all other frequencies and obscures low-power features in the time domain. Waveforms are therefore digitally notch- and high pass-filtered (see~\cite{proakis} for a discussion of each). The carrier is removed with an adaptive two-tap, recursive least squares (RLS) filter~\cite{dsp_rls_notch}. The high pass (HP) filter passes all frequencies above some cut-off frequency. It is designed using the Parks-McClelland~\cite{dsp_parks_mcclelland} algorithm with input zero amplitude filter response at 31.25 MHz and 43.75 MHz. Here it is referred to simply as the 30~MHz HP filter. Figure~\ref{fig:snapevdisp} shows the first 10~$\mu$s of the 132~$\mu$s snapshot event display after applying a 54.1~MHz notch and a 30~MHz HP filter. Near the beginning of the time and frequency domain plots the high amplitude 54.1~MHz frequency component can be seen. It diminishes as the notch filter adapts its phase and amplitude to cancel the carrier.

\begin{figure}[!h]
\centerline{\includegraphics[trim=0.0in 0.0in 0.0in 0.0in, clip=true, width=0.5\textwidth]{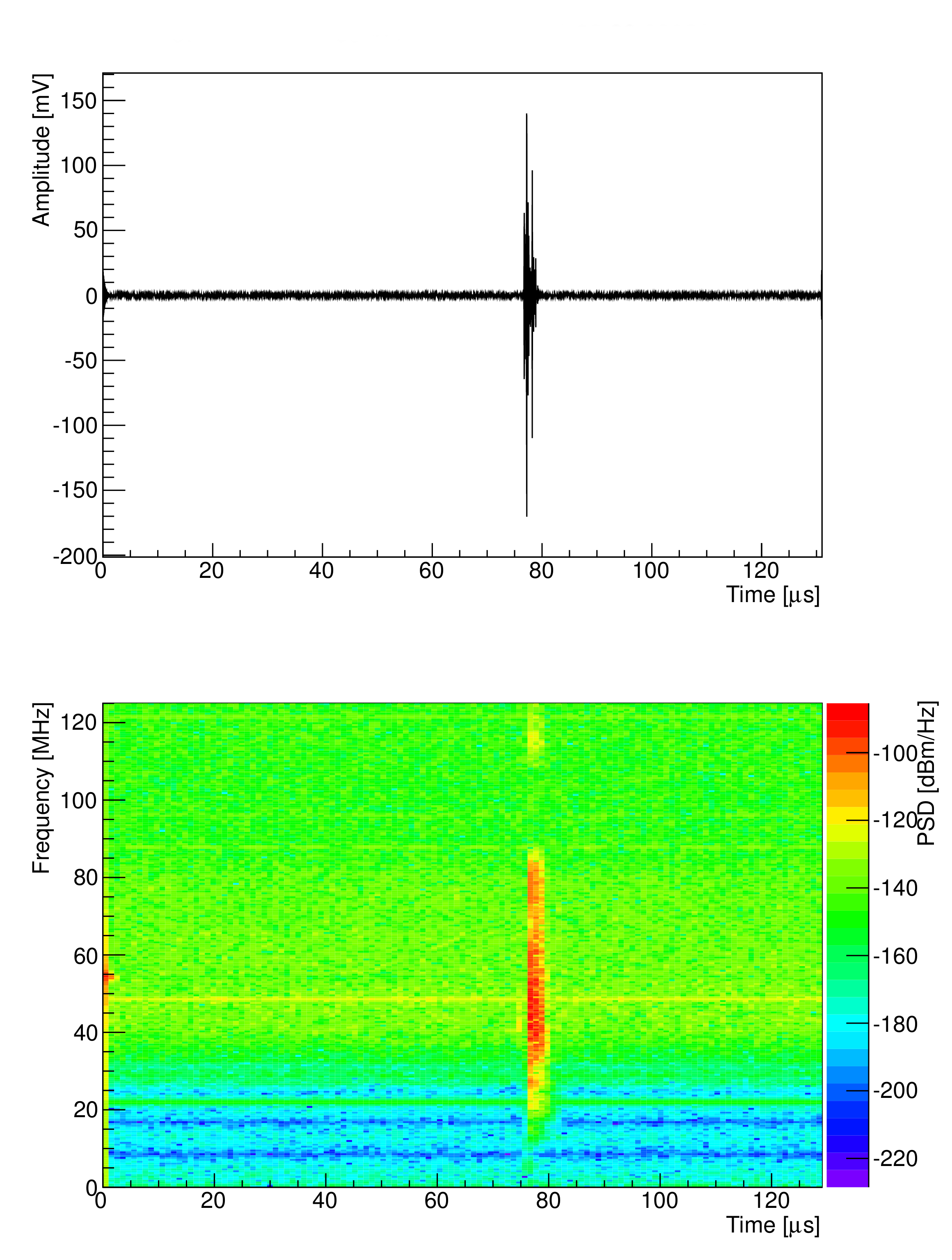}}
{\caption{Event display of high amplitude, transient broadband noise. This type of noise has high MF response.}
\label{fig:transient}}
\end{figure}

Radio noise is minimal at the location of the receiver. There are no static noise sources, but intermittent carriers and spurious impulsive noise must be considered in the background estimation. Broadband transients (Figure~\ref{fig:transient}) are the most common noise source and pose the worst threat of interfering in analysis because of the similarity to radar echoes, being broadband with short duration. Detection sensitivity is ultimately limited by the DAQ noise floor which has been shown to be consistent with galactic radio noise~\cite{taranim}.

\section{Signal Search}
\label{sec:sigsearch}
Section~\ref{sec:scattering} describes the thin-wire approximation to the RCS. It is emphasized that this model is a reasonable representation of the scattering geometry of the true RCS and that it overestimates the received power because collisional damping and skin depth are not included. Any observed signals are expected to have absolute power significantly below the simulation prediction and the same frequency versus time characteristic. 

If signal power were not overestimated, receiver data would be replete with radar echo coincidences with TA reconstructed events. The following analysis shows that this is not the case. Data are searched with a matched filter (MF) tuned to a specific event via echo simulation. Section~\ref{sub:mf} introduces the concept of matched filtering.  Section~\ref{sub:front_sim} gives a description of how the RF front end is simulated such that simulated echoes have the same frequency characteristics as the physical RF front end. In Section~\ref{sub:mf_thresh}, the MF threshold is described and the following two sections detail the signal search. 

\subsection{Matched Filter}
\label{sub:mf}
Matched Filtering (MF) is a digital signal processing (DSP) technique for detecting the presence of a known signal or ``template'' in a test waveform. The MF template is the ideal (no noise) digitized signal. The MF response is the convolution between the test waveform and the time-reversed template. 

\begin{figure*}[t]
\centerline{\mbox{\includegraphics[trim=0cm 0cm 0cm 0cm,clip=true,width=0.70\textwidth]{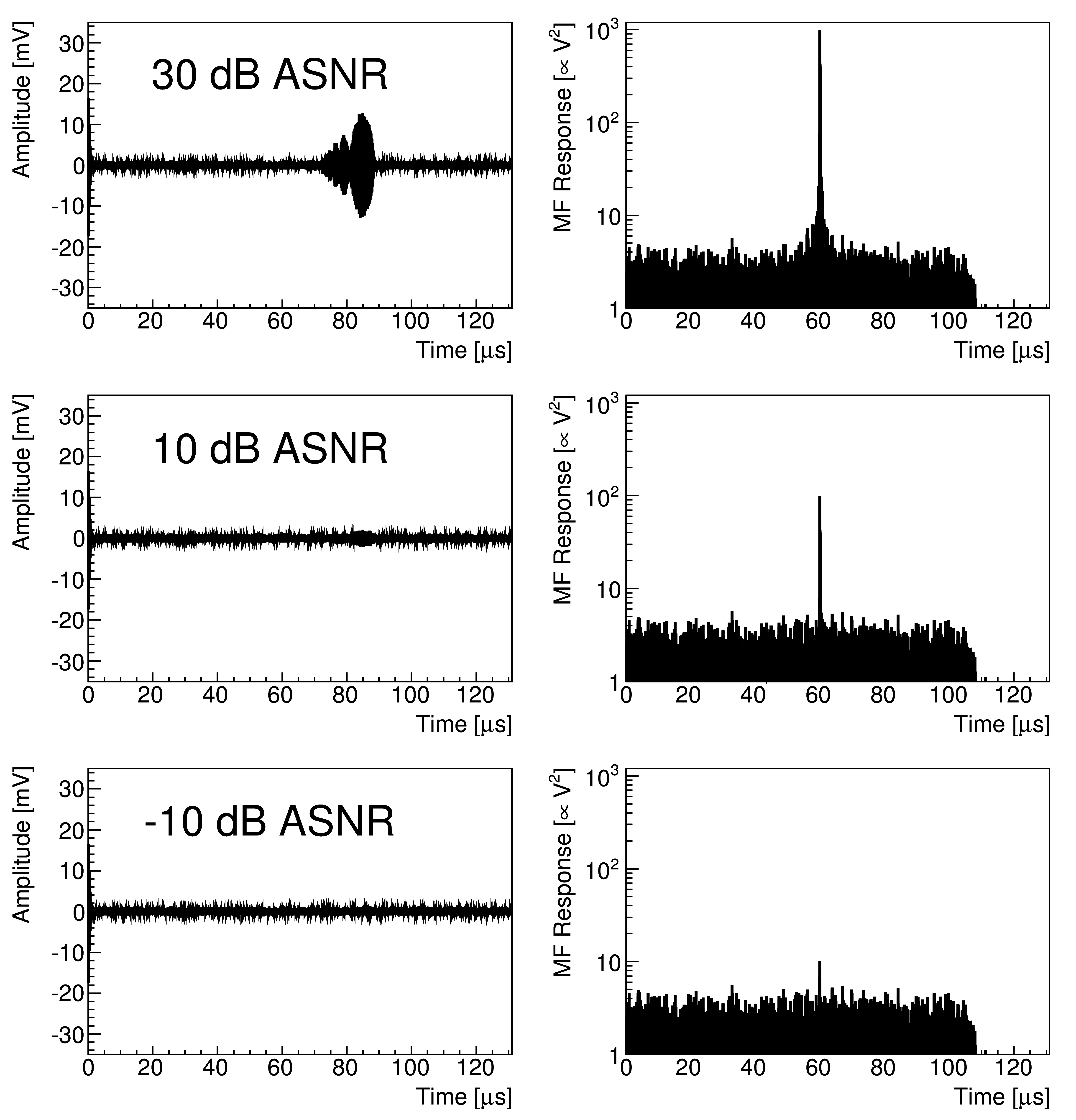}}}
\caption{Three examples of test waveform composed of a notch and HP filtered snapshot background superimposed with a simulated canonical radar echo (left plots) and their Matched Filter (MF) responses (right plots). The MF response is calculated using same superimposed simulated echo waveform as a template. Peak response occurs at 60~$\mu$s at the location of the below-noise beginning of the superimposed waveform, where the template and superimposed echo are aligned and the dot product between the two is maximized. The top, middle and bottom plots show results from superimposing the simulated echo at 30, 10 and $-10$~dB amplitude signal-to-noise ratio (ASNR). 
\label{fig:mf_example}}
\end{figure*}

Figure~\ref{fig:mf_example} shows three examples of a test waveform and MF output. In each case a simulated radar echo is superimposed on a random, filtered snapshot waveform. A template identical to the superimposed wave is used to calculate the absolute value of MF response as a function of the time after the convolution calculation begins (right plots). Peak MF response occurs at 60~$\mu$s where the template and the superimposed wave coincide in time and phase. The amplitude signal-to-noise ratios (ASNR)
\begin{equation}
\label{eq:asnr}
\text{ASNR} = \frac{V_{\text{max, c}}^2}{\sigma_{\text{noise}}^2}\,.
\end{equation}
(where $V_{\text{max, c}}$ is the maximum chirp amplitude and $\sigma_{\text{noise}}$ is the standard deviation of the background noise) of the three examples are 10, 0 and $-10$~dB. 
In this analysis, noise comprises all frequencies that remain after notch and HP filtering. (The more common SNR is defined as 
\begin{equation}
\label{eq:snr}
\text{SNR}=\frac{P_{\text{c}}}{\sigma_{\text{noise}}^2}\,,
\end{equation}
where $P_{\text{c}}$ is the chirp signal power, the square of the chirp waveform RMS.)

The simulated radar echoes are decimated down to the DAQ rate of 250~MS/s. Templates are then scaled such that the maximum voltage sample is 1.0~mV. The MF response is calculated in the trigger range of interest by computing the inner product of the template and predicted waveform. The result is a series of MF responses as a function of time in the window when the MF is applied, similar to the MF responses in Figure~\ref{fig:mf_example}. Only the maximum value (peak response) from the series of MF outputs is saved for further analysis. 

\subsection{Front End Simulation}
\label{sub:front_sim}
\begin{figure}[!h]
\centerline{\mbox{\includegraphics[trim=0cm 0cm 0cm 0cm,clip=true,width=0.5\textwidth]{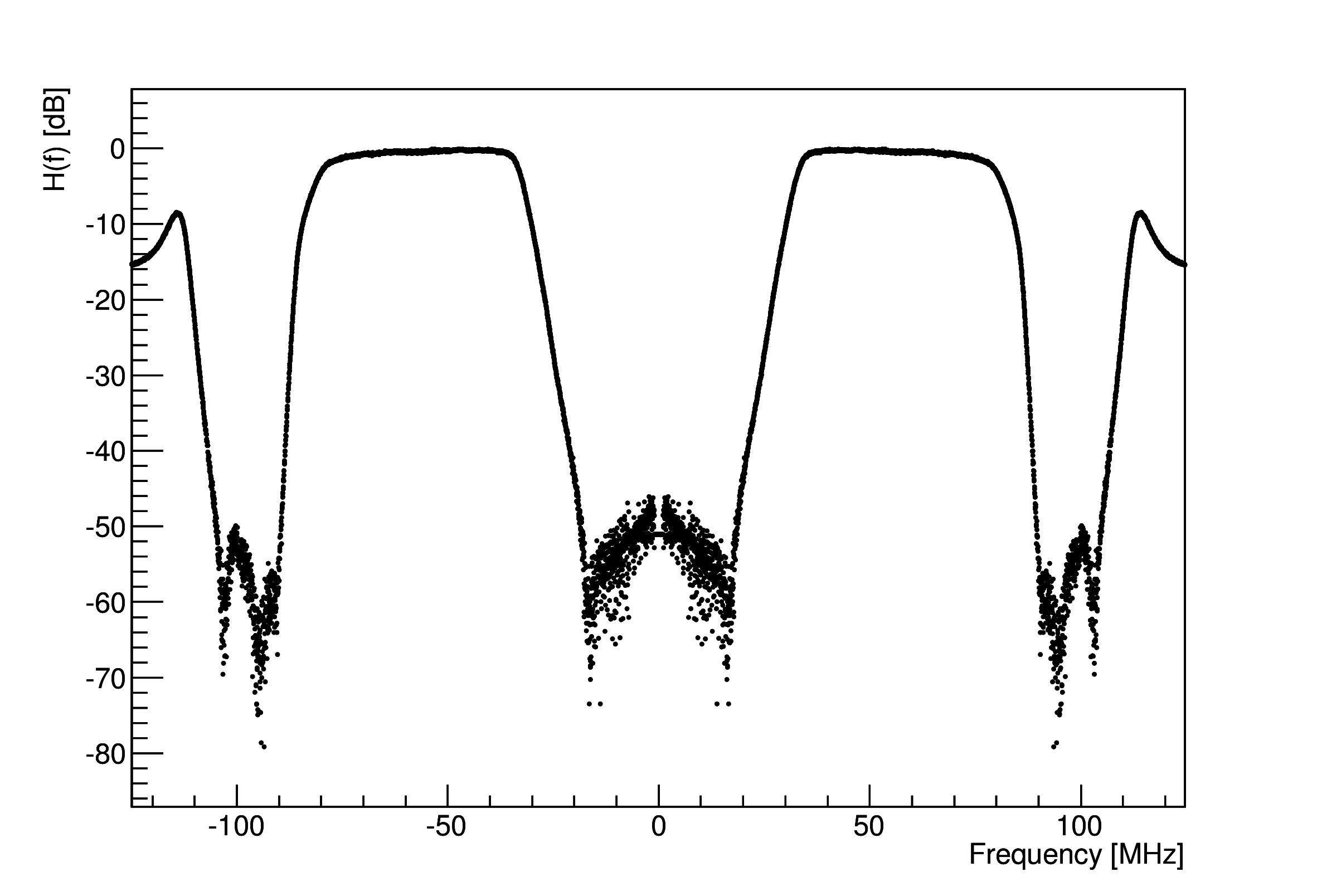}}}
\caption{Frequency response $H$, which has been reflected about 0~MHz, or DC. These data are the transmission coefficient $S_{21}$ which is measured at filter bank 3 and used to emulate a digital filter. 
\label{fig:freq_resp}}
\end{figure}

Simulation of the receiver front end electronics in TARA is necessary to understand the response of the system to theoretical chirp signals and, ultimately, to estimate the cross-section upper limit of UHECR air showers. The receiver antenna response is included in the simulation when radiation pattern lookup tables are queried for the gain specific to a shower segment, the location of which has specific geometry in the antenna coordinate system. Other front end components are characterized through the frequency response $H$ (see Figure~\ref{fig:freq_resp}), measured as the transmission coefficient ${\rm S}_{21}$ {\em in situ} with a two-port vector network analyzer (VNA). Components in the RF frontend including RF limiter, amplifiers and filters, are detached from the antenna and transmission line to the DAQ. The input and output of the RF chain is then connected to the VNA, where $S_{21}$ is measured at discrete frequency points in the $0 - 125$~MHz DAQ band.

Data from the ${\rm S}_{21}$ measurement are used to construct a filter which can be applied in the time domain to simulated echoes. Echoes filtered in this manner exhibit a frequency response as if the waveform had passed through the actual RF frontend. The procedure for creating the time domain filter from desired frequency response $H$ is introduced in digital signal processing (DSP) texts~\cite{proakis}. In practice, filter coefficients are obtained by calculating the inverse Fourier transform of $H$.

\begin{figure}[h]
\centering
\begin{subfigure}{0.5\textwidth}
\includegraphics[trim=0.0cm 0.0cm 0.0cm 0.0cm,clip=true,width=1.0\linewidth]{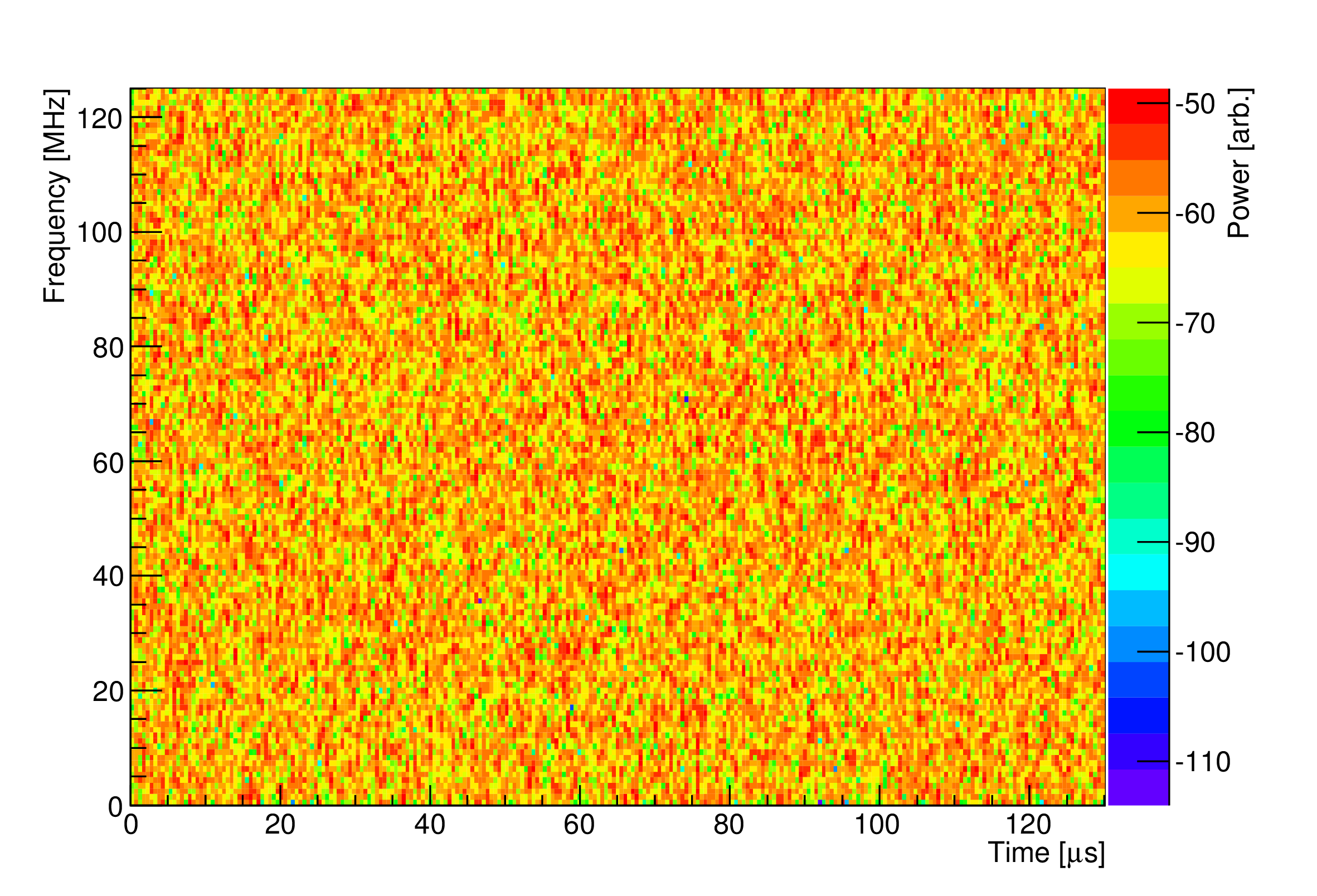}
\end{subfigure}
\begin{subfigure}{0.5\textwidth}
\includegraphics[trim=0.0cm 0.0cm 0.0cm 0.8cm,clip=true,width=1.0\linewidth]{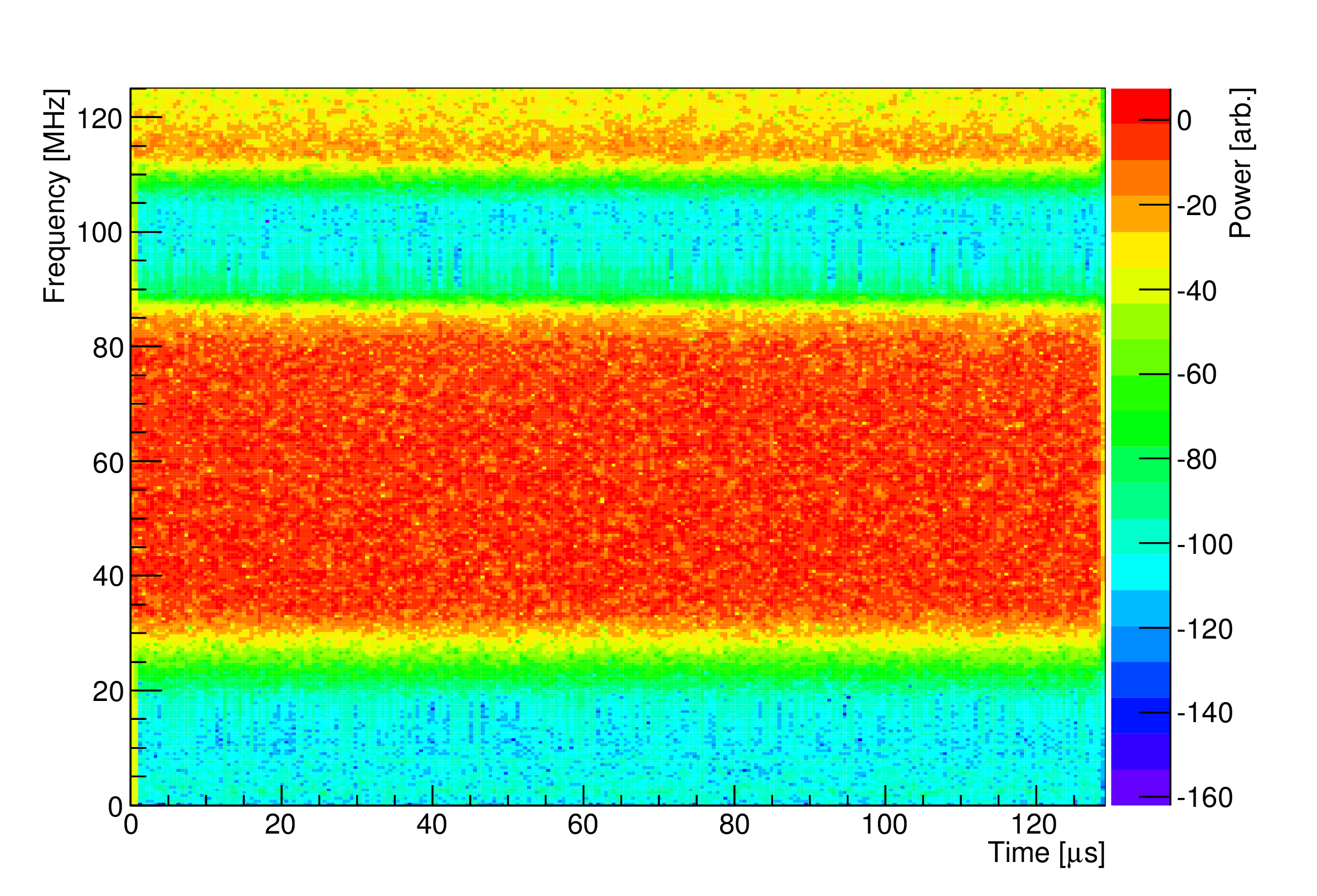}
\end{subfigure}
\caption{Spectrograms showing simulated Gaussian noise (top) and the same noise filtered with an FIR filter designed from the frequency response shown in Figure~\ref{fig:freq_resp}. Amplitude features can be directly compared between the spectrogram and the desired response.
\label{fig:noise_filter_spectrograms}}
\end{figure}

Figure~\ref{fig:noise_filter_spectrograms} shows two spectrograms featuring Gaussian and post-filtered Gaussian noise. The frequency response $H$ of the RF frontend for the acquisition channel used in the present analysis is plotted in Figure~\ref{fig:freq_resp}, where it has been reflected about zero. Features of the frequency response are clearly seen in the bottom plot in Figure~\ref{fig:noise_filter_spectrograms}. All simulated echoes are filtered in this manner before being used in analysis.

\subsection{MF Threshold}
\label{sub:mf_thresh}
Snapshots of the TARA data stream are selected from the FD run according to the following criteria: the transmitter must be on at the time the snapshot was acquired, the snapshot must occur in a five-minute bin in which the average FD trigger rate is at least 1~Hz, and no dead channels or DAQ interruptions can be present in the waveform. The same quality cuts are applied to matched FD triggers. 

Parameters from matched TA events are used to simulate radar echoes. Each matched TARA FD-trigger has a unique, simulated echo which is used to create a MF template. The radar echo simulation produces theoretical received signals according to transmitter power, detector and shower geometry, antenna gain and shower parameters. Matched triggers have associated CR energy, geometry, \xmax\ and core locations determined from the reconstructed FD event to which they are matched. These data are used in the radar echo simulation. 

Templates are created by filtering simulated echoes to emulate the RF front end response (see Section~\ref{sub:front_sim}), decimating to 250~MS/s, and then truncating by removing samples at the beginning and end of the waveform that are below 5\% of the peak. Low amplitude parts of the template do not greatly affect MF response and processing time is reduced by removing low amplitude tails. 

\begin{figure}[h]
\centerline{
\includegraphics[trim=0.0cm 0.0cm 0.0cm 0.0cm,clip=true,width=0.5\textwidth]{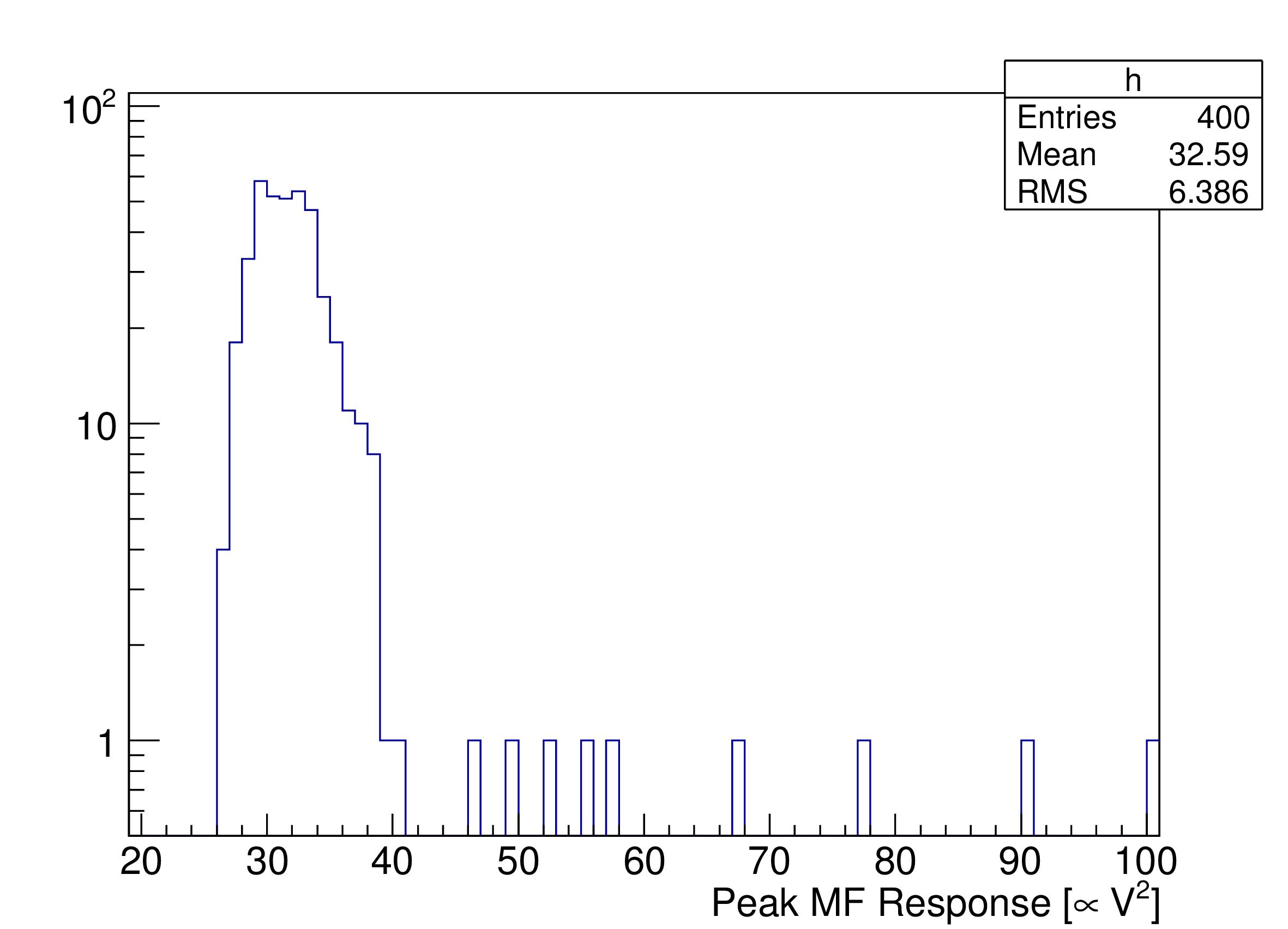}}
\caption{Peak MF response distribution for an August, 2013 FD-triggered matched event using 400 snapshots with a simulated radar echo MF template. The 3~RMS threshold is 51.8.} 
\label{fig:mfdist}
\end{figure}

It is necessary to estimate the noise response of each MF independently. Therefore, a peak MF response distribution is created for each matched FD-triggered event by applying the simulated echo template as a MF to a set of 400 selected snapshots, which are first notch- and then 30~MHz HP-filtered. The distribution is used to calculate a MF threshold at the 3~RMS (three standard deviations) level. An example distribution and threshold is shown in Figure~\ref{fig:mfdist}.

\begin{figure}[!h]
\centerline{\mbox{\includegraphics[trim=0cm 0cm 0cm 0cm,clip=true,width=0.5\textwidth]{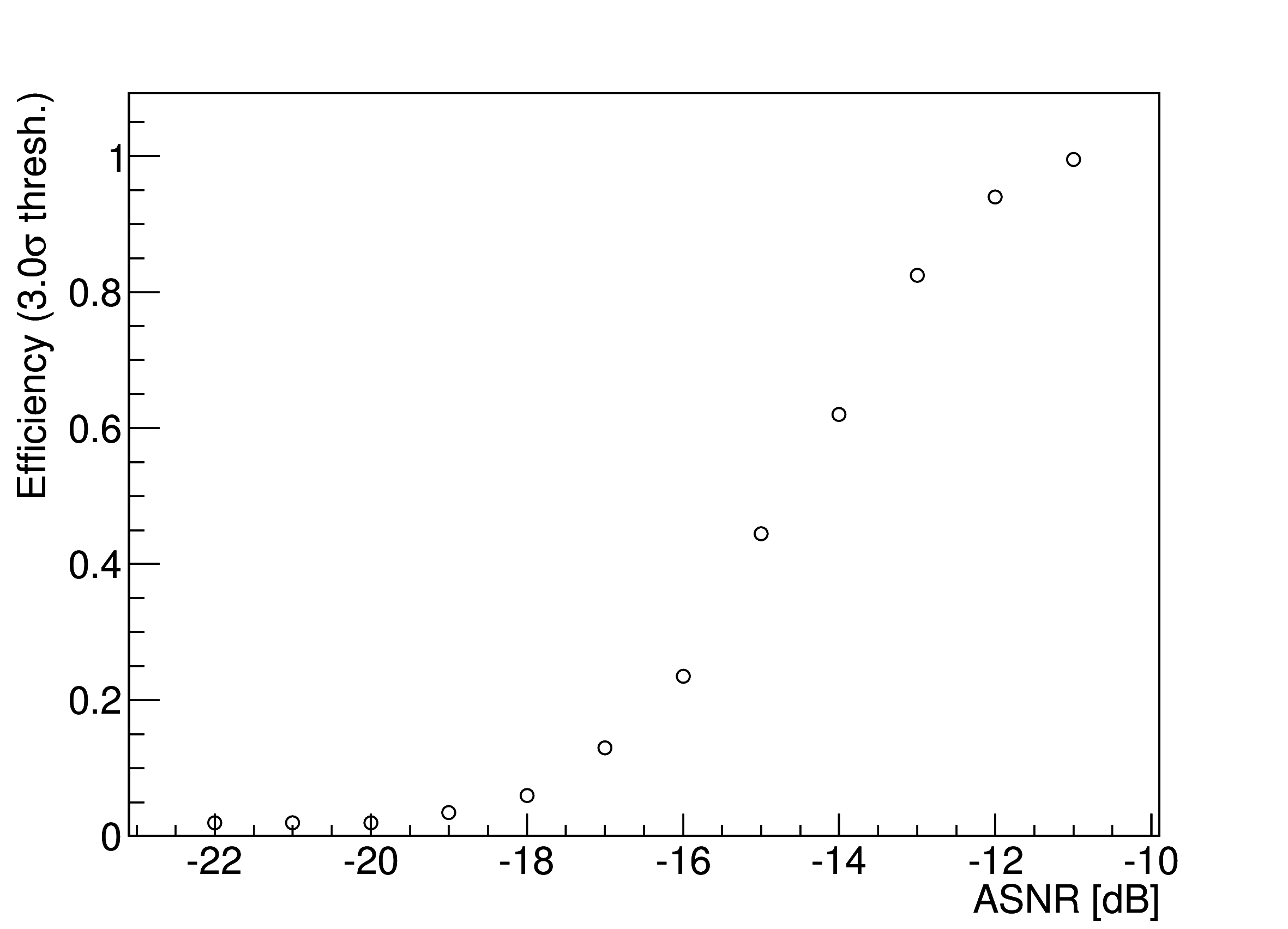}}}
\caption{Matched filter detection efficiency as a function of ASNR using a simulated radar echo, for a 10~EeV shower midway between transmitter and receiver. Selected snapshots from the August, 2013 FD run are used both to determine the 3~RMS response threshold and as backgrounds on which scaled echo waveforms are superimposed.
\label{fig:detection_performance}}
\end{figure}

This MF detection scheme, in which the MF peak response distribution mean and RMS determine the threshold, has high efficiency at signal levels more than 10~dB ASNR below the galactic noise floor. Figure~\ref{fig:detection_performance} shows detection efficiency as a function of ASNR for a typical event geometry. 

\subsection{Search: Reconstructed Events}
\label{sub:posdetection} 
Of 1,206 events processed in the data set (including FD runs from August, 2013 to April, 2014), 17 events are found to have MF response greater than 
3~RMS. These are considered as possible positive detections of scattered radar waves from EAS. The statistical significance of this value is calculated using the expected value of snapshot false positives.

Recall that distributions of the snapshot peak MF response are used to calculate the detection threshold. 
In the majority of these distributions, which are created for each event, several snapshot MF responses exceed the threshold. Figure~\ref{fig:mfdist} shows the MF response distribution for an event from August, 2013. The threshold is $\bar{x} + 3\text{RMS} = 51.8$. Seven snapshots out of 400 exceed the threshold. 

\begin{figure}[t]
\centerline{
\includegraphics[trim=0.0cm 0.0cm 0.0cm 0.0cm,clip=true,width=0.5\textwidth]{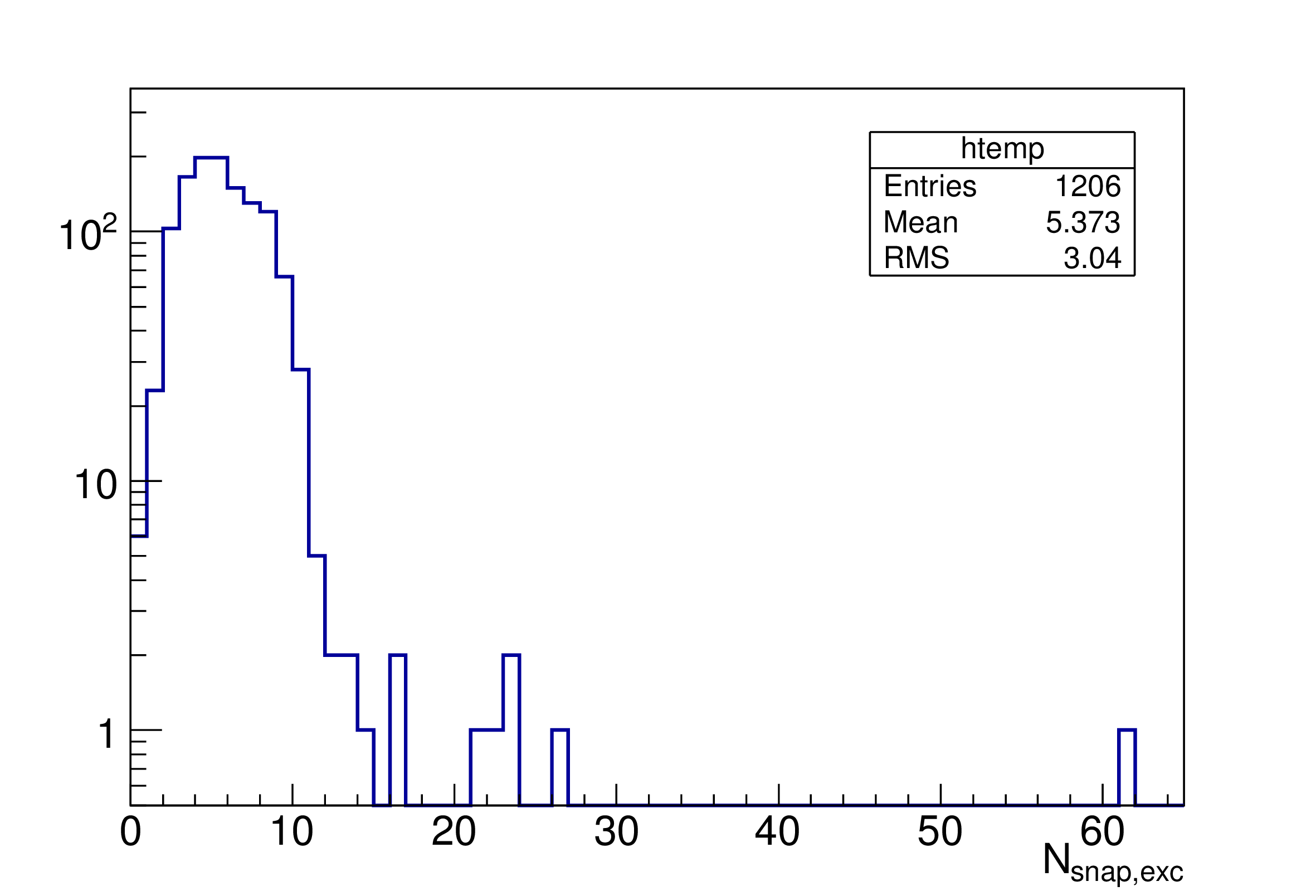}}
\caption{Distribution of $N_{\text{snap,exc}}$, the number of snapshots with peak MF responses that exceed threshold. One entry per matched FD-triggered event.}
\label{fig:nexc}
\end{figure}

Figure~\ref{fig:nexc} shows the 
number of snapshots exceeding threshold for all 1,206 events included in the analysis. On average, 1.34\% of snapshots exceed the 3~RMS threshold, leading to a false-positive estimate of 16.2 events for our current sample. This is completely consistent with our observation of 17 positive detections. 
We thus interpret our positive detections as false positives, caused by common background noise.

\subsection{Search: Nearby Events}
\label{sub:nearby}
The previous analysis only considers 
events which can be fully reconstructed by the FD. This restricts the number of events by requiring a minimum $R_p$ of 1~km (see Table~\ref{tab:fdcuts}). Events that occur close to the FD pass through the telescope field of view quickly relative to the DAQ trigger sampling rate such that shower evolution and profile are not clearly observed: the events are not reconstructible. The bistatic radar equation (Equation~\ref{eq:bi}) predicts maximum received power when the target is close to either the transmitter or receiver. There are no FD-triggered events that occur near the transmitter station because it is beyond the detection range of the FD, but many events occur close to the FD where the TARA receiver is located.

A search was conducted for evidence of CR radar echoes in FD triggers that fail the $R_p$ cut. After time-matching TARA FD triggers with events that fail the $R_p$ cut in the same manner as above, 2124 events remain. Simulated radar echoes of nearby Monte Carlo events are used to create a set of matched filters which will be used in the search. A large set of FD-triggering Monte Carlo events was narrowed down to a total of 673 events by requiring that $E_0 > 3\times 10^{17}$~eV and $R_{p} < 1$~km. 

\begin{figure}[t]
\centering
\begin{minipage}{0.48\textwidth}
\centering
\includegraphics[trim=0.0cm 0.0cm 0.0cm 0.0cm,clip=true,width=1.0\textwidth]{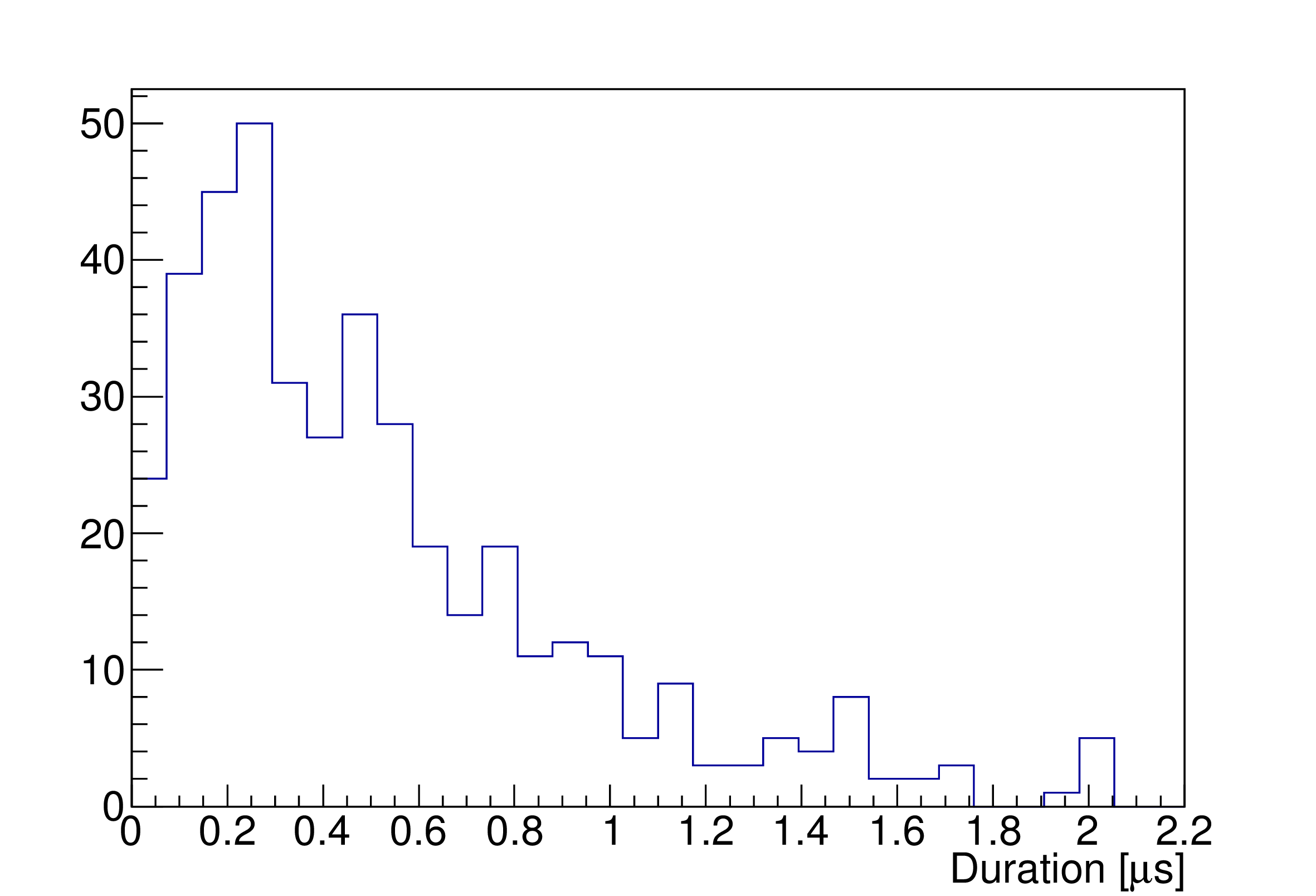}
{\caption{Distribution of radar echo durations within the TARA [40, 80]~MHz passband. Duration is defined as the period during which the peak power occurs, with endpoints 10~dB below the peak.}
\label{fig:dur_dist}}
\end{minipage}\hfill
\begin{minipage}{0.48\textwidth}
\centering
\includegraphics[trim=0.0in 0.0in 0.0in 0.2in, clip=true, width=1.0\textwidth]{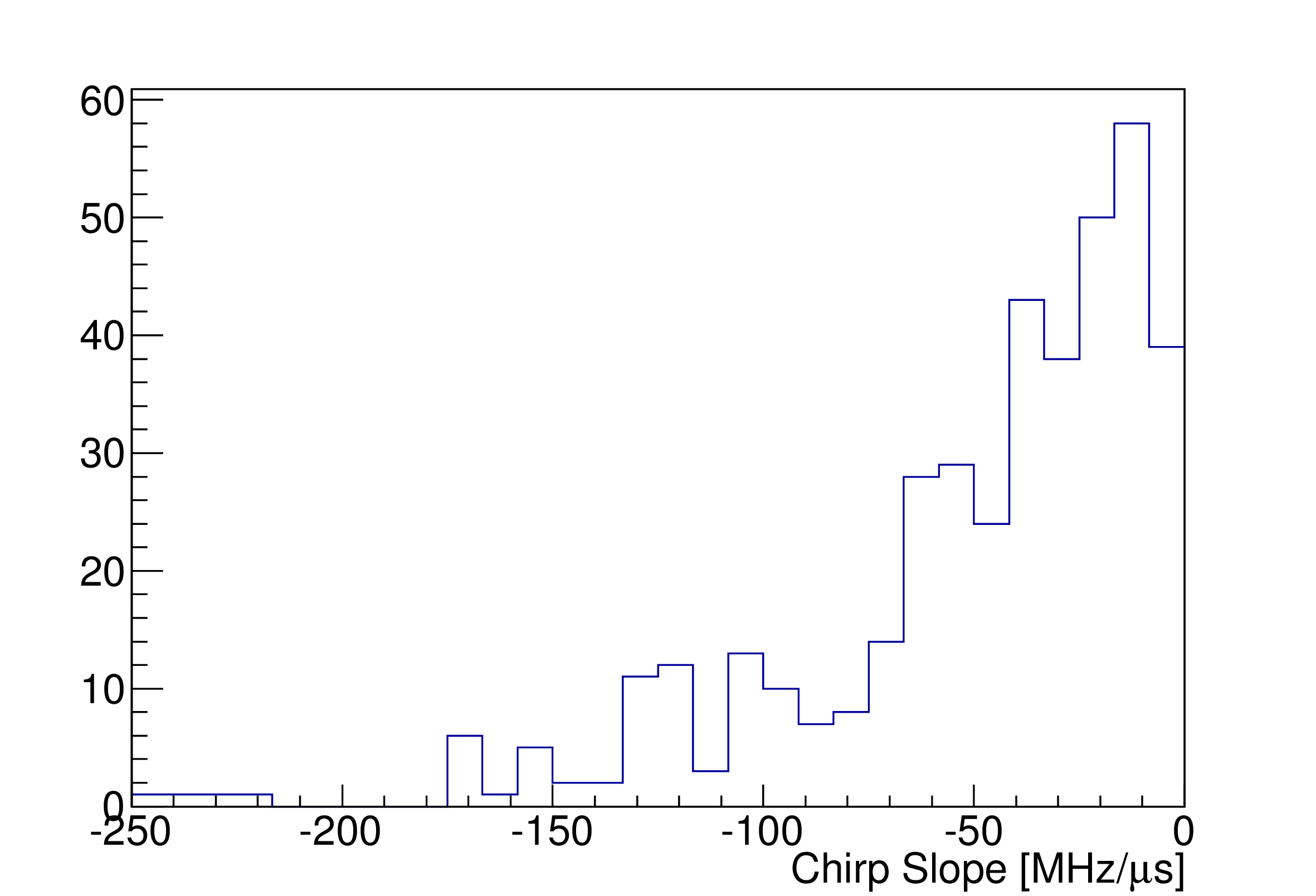}
{\caption{Distribution of radar echo chirp slopes within the TARA [40, 80]~MHz passband. The slope is calculated from a simple straight line fit to the power spectrogram.}
\label{fig:slope_dist}}
\end{minipage}
\end{figure}

Radar echoes from the nearby and unreconstructible Monte Carlo events are simulated and analyzed for chirp slope and duration within the passband. Due to the proximity of the shower and receiver, many of the simulated events are undetectable due to large frequency shifts that place echoes above the 80~MHz passband ceiling. Duration and slope distributions of the simulated events are shown in Figures~\ref{fig:dur_dist} and~\ref{fig:slope_dist}. Entries with zero duration or slope are events with echoes outside the passband.

A set of 21 linear chirp, constant amplitude matched filters (MF) were created to broadly represent the Monte Carlo distributions: MFs are 400~ns in duration, span 140~MHz of chirp rate in seven discrete steps (10, 20, 30, 40, 50, 100, and 150~MHz/$\mu s$), and represent three different frequency bands, starting at 80, 70, and 60~MHz. A 3RMS threshold is determined for each of the 21 MF templates. No signal candidates were found in the relaxed $R_{p}$ signal search.

\section{RCS Calculation}
\label{sec:rcscalc}
\subsection{$\Gamma_{90}$ Calculation}
\label{sub:gamma90}
The search described in the previous section indicates that there is no echo signal present in the dataset described in this paper. Now, we use this information to place quantitative limits on the RCS for particular air shower events. We make use of the thin-wire model described in Section~\ref{sec:scattering}, and assume that the cross section is proportional to the thin wire cross section $\sigma_{\text{TW}}$:
\begin{equation}
\sigma_{\text{EAS}} = \Gamma \sigma_{\text{TW}}\,,
\end{equation}
(where $\Gamma$ is a dimensionless scale factor) and place 90\%~c.l. upper limits on the RCS within the thin-wire model. The basic strategy is to vary the scale factor $\Gamma$ until 90\% of the simulated waveforms exceed the MF threshold determined by the algorithm described in Section~\ref{sub:mf_thresh}. This value of $\Gamma$ is then designated $\Gamma_{90}$ for a particular event.

\subsection{$\Gamma_{90}$ Results}
\label{sub:indsf}

\begin{figure}[h]
\centerline{
\includegraphics[trim=0.0cm 0.0cm 0.0cm 0.0cm,clip=true,width=0.5\textwidth]{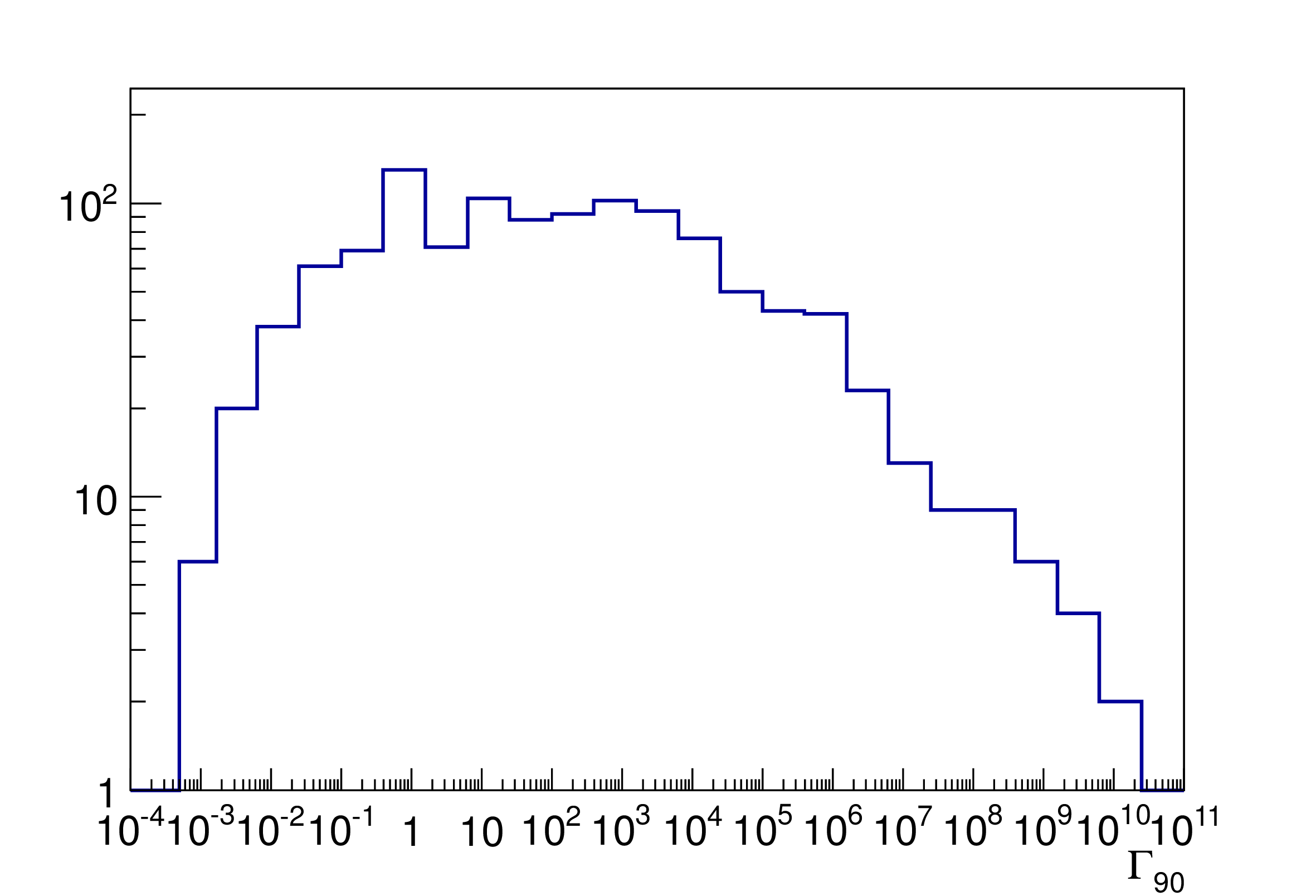}}
\caption{$\Gamma_{90}$ distribution of all negative detection events. Large $\Gamma_{90}$ occurs when matched-event geometry specifies an EAS that occurs outside the antenna main lobe.}
\label{fig:sfhisto}
\end{figure}

Figure~\ref{fig:sfhisto} shows a histogram of all $\Gamma_{90}$ values for events with a MF response of less than 3~RMS (``negative detection''). 

\begin{figure}[h]
\centerline{
\includegraphics[trim=0.0cm 0.0cm 0.0cm 0.0cm,clip=true,width=0.5\textwidth]{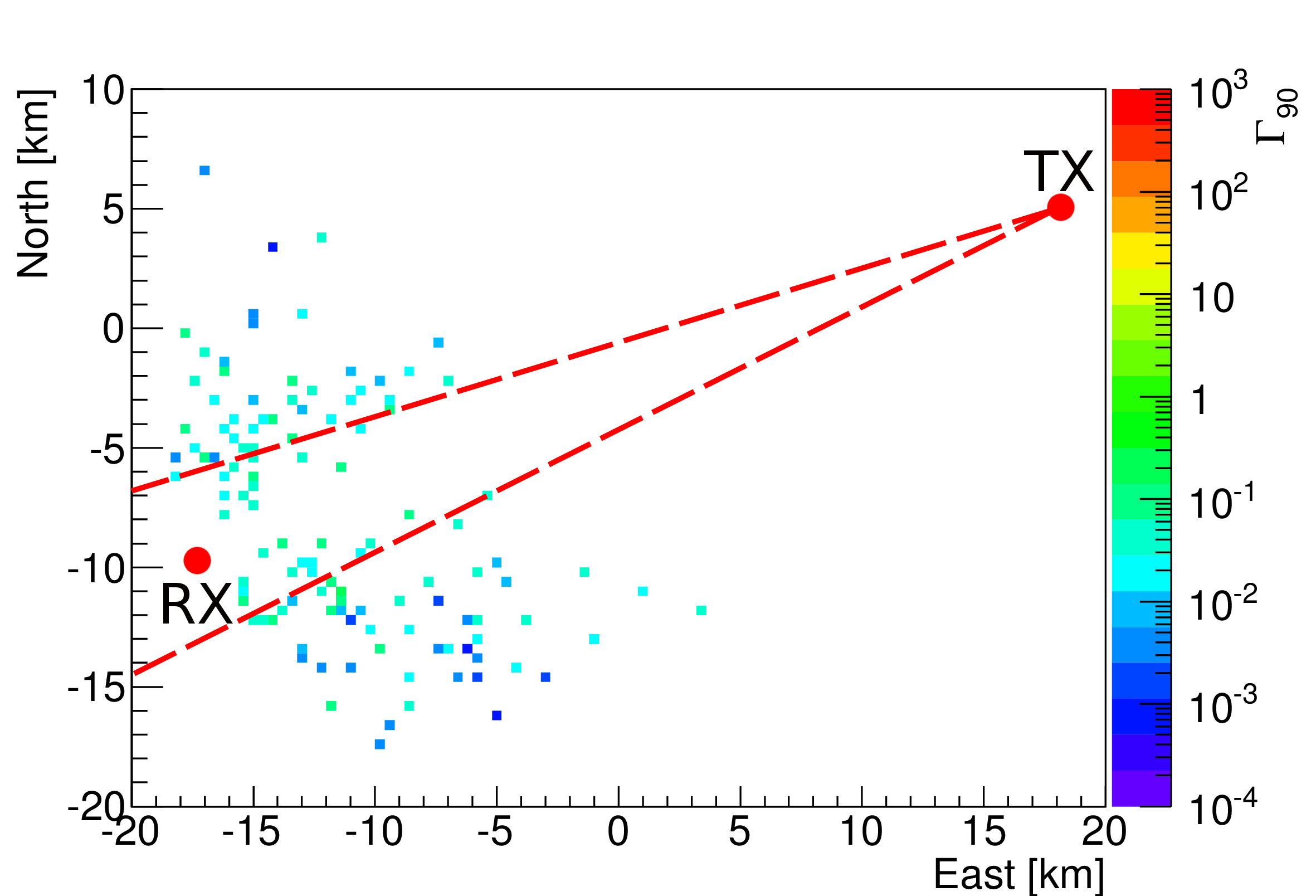}}
\caption{$\Gamma_{90}$ (color scale) for negative detection events with $\Gamma_{90} < 0.1$ shown at reconstructed core locations in Telescope Array CLF coordinates. Red dashed lines mark the primary beam -3~dB beamwidth.}
\label{fig:sfcoreloc_lowsf}
\end{figure}

Figure~\ref{fig:sfcoreloc_lowsf} is a plot showing the locations at which the air shower core strikes the ground, including only events with $\Gamma_{90} < 0.1$. Red, dashed lines indicate the transmitter antenna $-3$~dB beamwidth. Event core locations are clustered near -- but not under -- the main lobe. Very few of the selected events will have small zenith angles due to the horizontal polarization of the transmitted signal. Small scale factor events must be those that pass through the main lobe, therefore there will be few events along the center of the beam. $\Gamma_{90} > 0.1$ events comprise the majority of events and are those to which TARA is least sensitive.

\begin{figure}[h]
\centerline{
\includegraphics[trim=0.0cm 0.0cm 0.0cm 0.0cm,clip=true,width=0.5\textwidth]{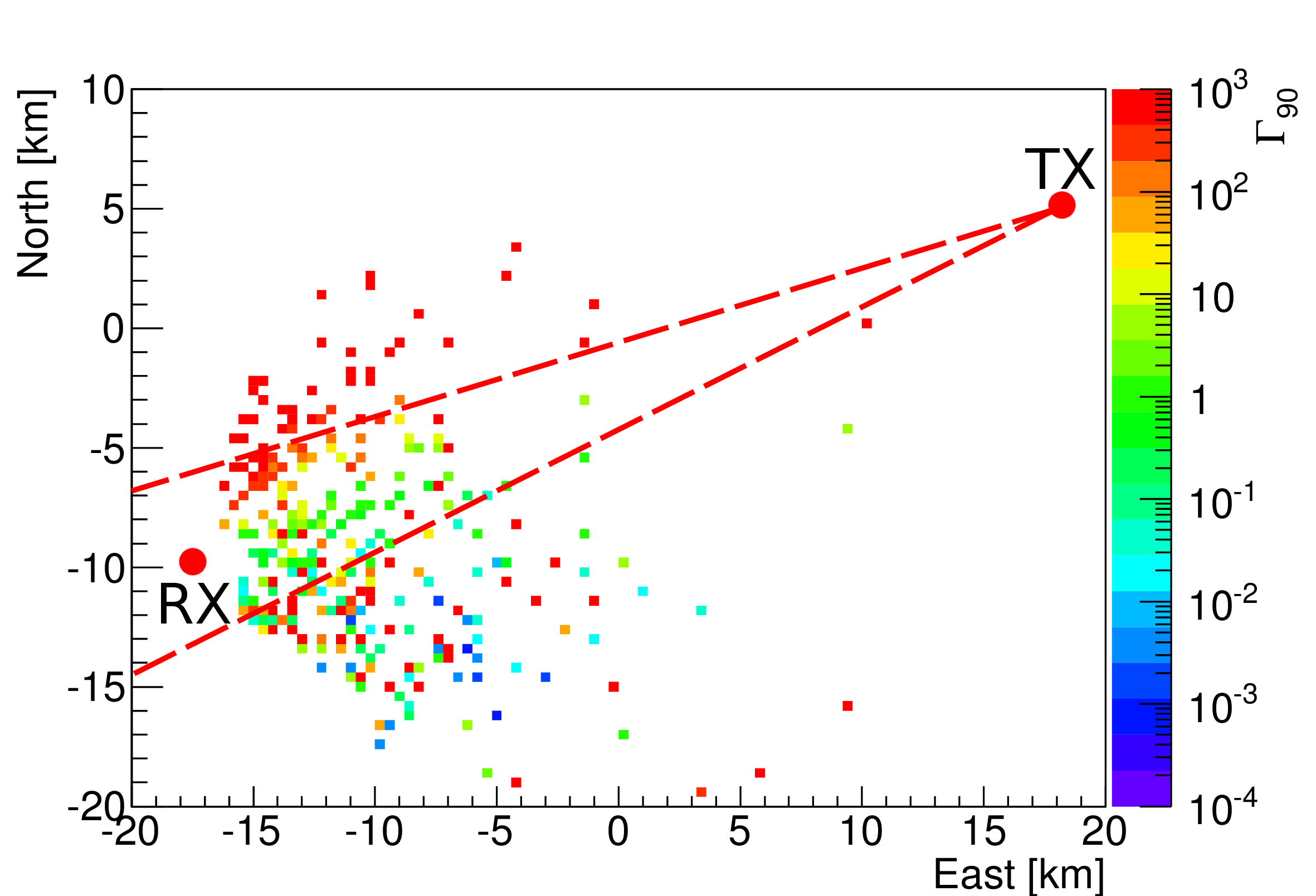}}
\caption{$\Gamma_{90}$ (color scale) negative detection events restricted to those with $90^{\circ} < \text{azimuth} < 180^{\circ}$ shown at reconstructed core locations in Telescope Array CLF coordinates. Red dashed lines mark the primary beam -3~dB beamwidth.}
\label{fig:sfcoreloc_narrowazi}
\end{figure}

The effect of the transmitter antenna main lobe on the scale factor can be seen in Figure~\ref{fig:sfcoreloc_narrowazi}, which restricts the set of events to those with azimuth greater than 90$^{\circ}$ and less than 180$^{\circ}$. (The azimuth origin is due east. All events in the plot point back to the source toward the second quadrant (upper left). Events above the beam have very high scale factors because those events do not interact with the main lobe. The lowest scale factor events are those that produce high echo voltages due to optimal geometry and interaction with the main lobe. Reversing the general pointing direction of events, by limiting azimuth to greater than 270$^{\circ}$ and less than 360$^{\circ}$, produces a similar plot, but with the high $\Gamma_{90}$ events below the main lobe and low $\Gamma_{90}$ above the main lobe. 

\begin{figure}[t]
\centerline{
\includegraphics[trim=0.0cm 0.0cm 0.0cm 0.0cm,clip=true,width=0.5\textwidth]{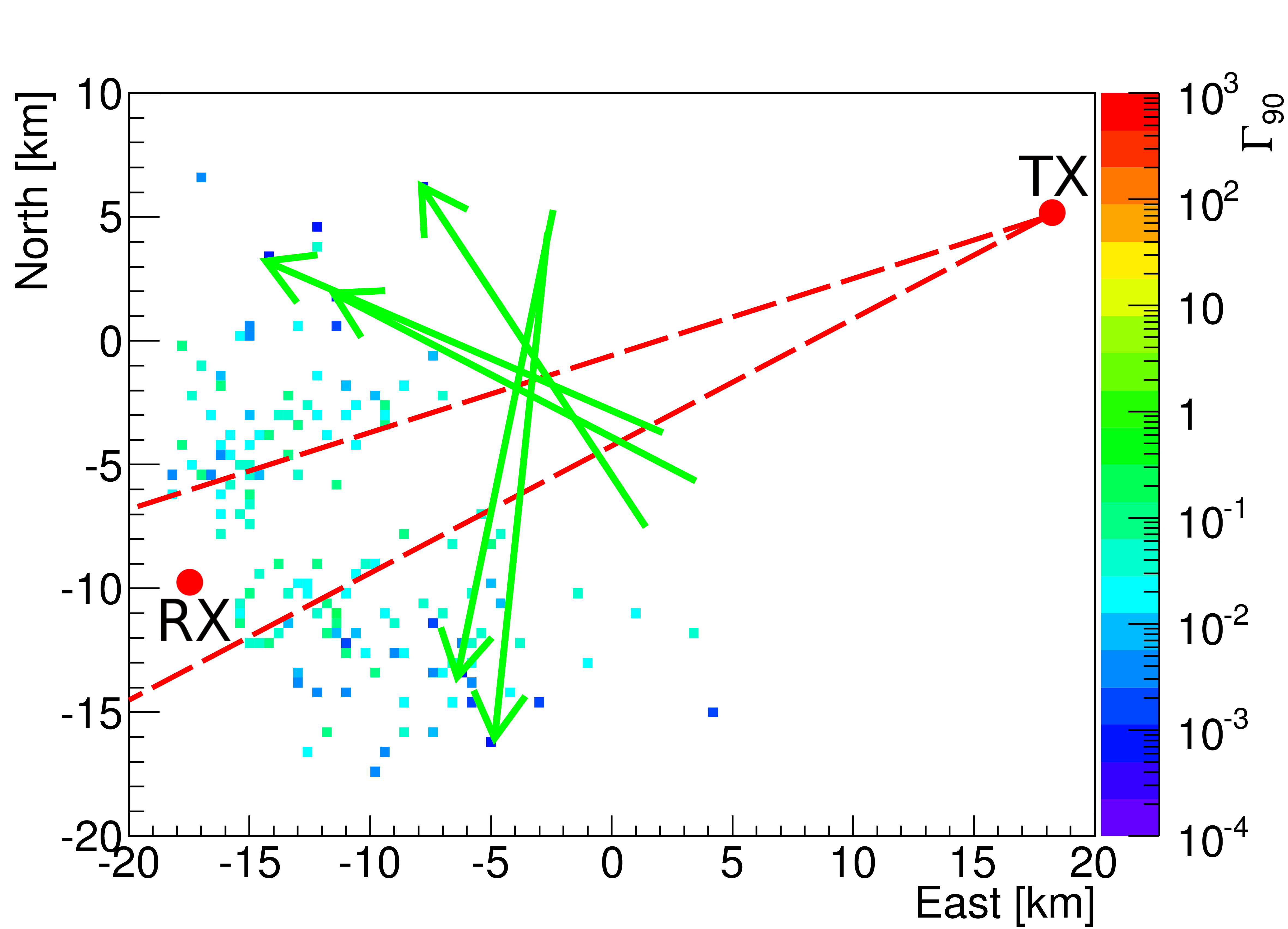}}
\caption{$\Gamma_{90}$ (color scale) for negative detection events with $\Gamma_{90} < 0.1$, similar to Figure~\ref{fig:sfcoreloc_lowsf}. Additionally, green arrows point in the direction the shower propagates through the atmosphere at the core location of the five lowest $\Gamma_{90}$ events. Arrow length is proportional to zenith angle.}
\label{fig:lowsf_coreloc_zenline}
\end{figure}

Figure~\ref{fig:lowsf_coreloc_zenline} is similar to the plot in Figure~\ref{fig:sfcoreloc_lowsf}, but includes green arrows that point at the core location of the five lowest $\Gamma_{90}$ events and point in the direction the shower travels toward the ground. Arrow length is proportional to zenith angle. One observes that the five events with the lowest $\Gamma_{90}$ are highly inclined to match transmitter and receiver polarization and have azimuth values which allow them to interact with the main lobe. 

\subsection{Systematic Uncertainties}
\label{sub:uncertainty}
In the previous section, we describe the calculation of 90\%~c.l. upper limits on the radar cross section for particular Telescope Array events. Systematic uncertainties in the terms in the bistatic radar equation (Equation~\ref{eq:bi}) make necessary relatively small adjustments to the upper limits reported. 

The total transmitter power is logged, and is known at any particular time to better than 1\%. Transmitter forward gain is modelled as a function of angle by Numerical Electromagnetics Code~\cite{nec} (NEC), which has been confirmed by direct measurements as an accurate representation of the relative gain to within 10\%~\cite{taranim}. 

The receiver gain is also modeled using NEC, at 54.1~MHz. An uncertainty arises from variations in the receiver gain as a function of frequency, which is not included in the waveform simulations. This effect is estimated to be 3\% or less for events in geometries to which TARA is most sensitive. 

Monocular FD reconstruction uncertainties of approximately 17\% in the energy of air showers and approximately 70~g/cm$^2$ in the depth of shower maximum enter in to the expectation for the target cross-section. Within the thin wire model the effect due to uncertainty in energy is weak, owing to the log-squared dependence of cross section on wire diameter (and therefore energy, Equation~\ref{eq:thinwire}). Shifting the location of shower maximum $X_{max}$ by 100~g/cm$^2$ is also found to be a small effect, shifting $\Gamma_{90}$ by less than 1\%.

Uncertainties in the transmitter-to-target ($R_T$) and target-to-receiver ($R_R$) distances come from uncertainty in the air shower geometries arising from monocular fluorescence reconstruction. The largest effect derives from the uncertainty in the shower-detector plane angle $\Psi$, shown in Figure~\ref{fig:fdgeo}. The uncertainty in $\Psi$ is typically $8^{\circ}$. Fluctuations of either sign in $\Psi$ are found to have the effect of increasing $\Gamma_{90}$ by as much as 60\%. This is by far the dominant contribution to systematic uncertainty in the cross-section upper limit. We account for this systematic by increasing the lower limits by the amounts shown in Table~\ref{tab:mod_gamma90}.

\begin{table*}[t]
\caption{Summary of five lowest $\Gamma_{90}$ events. The values presented for $\Gamma_{90}$ include an increase --- by as much as 60\% --- due to systematic uncertainties associated with uncertainties in the shower-detector plane angle. Core location pairs (x,y) are in units of kilometers relative to TA's central laser facility (CLF). In this coordinate system the transmitter is located at (17.9, 4.7) and the receiver site is located at (--18.4, --9.9).  
\label{tab:mod_gamma90}}
\begin{center}
\begin{tabular}{|l|c|c|c|c|c|c|}
\hline
Date     & Energy~[EeV] & Core Loc.~[km] & Zen.~[deg.] & Azi.~[deg.] & \xmax~g/cm$^2$ & $\Gamma_{90}\,\times 10^{4}$ \\ \hline
\hline
20130809 &  1.22        & (-11.5,1.9)    & 65.7        & 301.6       & 772            & 13.4    \\ \hline
20130816 &  1.43        &  (-7.9,6.2)    & 68.6        & 280.5       & 755            & 10.8   \\ \hline
20130926 &  1.38        & (-14.3,3.2)    & 54.9        & 299.5       & 837            & 12.5   \\ \hline
20131105 &  1.83        & (-4.8,-16.0    & 59.6        & 121.4       & 805            & 13.1  \\ \hline
20131202 & 11.04        & (-6.4,-13.6)   & 62.7        & 114.6       & 859            &  7.7 \\ \hline
\end{tabular}
\end{center}
\end{table*} 

\subsection{Discussion; RCS Upper Limits}
\label{sub:rcs_upper}
\begin{figure}[h]
\centerline{
\includegraphics[trim=0.0cm 0.0cm 0.0cm 0.0cm,clip=true,width=0.5\textwidth]{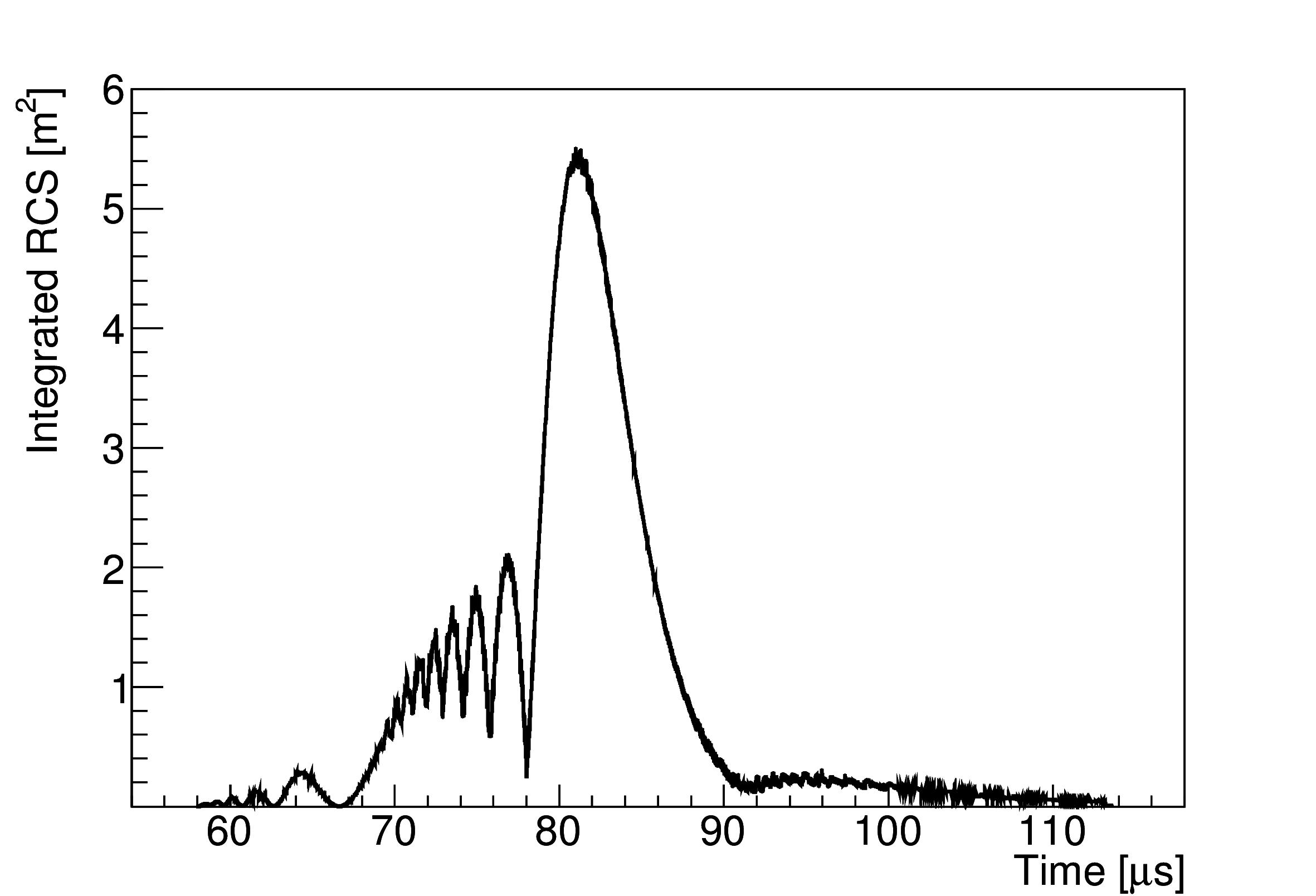}}
\caption{Integrated EAS RCS using the thin-wire approximation to RCS for a radar echo simulated with reconstructed shower parameters. Phase factors are included in the sum of the total RCS to properly account for each longitudinal shower segment. }
\label{fig:lowsf_ind5_intRCSvTime}
\end{figure}

Estimates of the detectable RCS for the TARA detector have been made previously~\cite{taranim}, and shown to be of order 50~cm$^2$, without the post-processing technique used in the present paper. For comparison, we may consider the best result in Table~\ref{tab:mod_gamma90}:
\begin{equation}
\sigma \leq 0.00077~\sigma_{\text{TW}} \hspace{0.3cm} {\rm (90\%~c.l.)} 
\end{equation}
Figure~\ref{fig:lowsf_ind5_intRCSvTime} shows the total integrated thin-wire RCS as a function of time during shower evolution and propagation toward the ground for this specific event. The RCS is integrated over every longitudinal segment of the shower with plasma age less than 5~$\tau$ at each 2~ns time step and phase factors attributed to differing path length are included in each term in the sum. The peak RCS is 5.5~m$^2$.

After including detector sensitivity, nonobservation of signal and the effect of systematic uncertainty the effective RCS upper limit may be expressed as a product of the peak RCS with $\Gamma_{90}$, equal to approximately 42~cm$^2$ at 90\%~c.l. This is in good agreement with the earlier estimate. The implications of such a small upper limit will be discussed in the conclusion. 

\begin{figure}[h]
\centerline{\includegraphics[trim=0.0in 0.0in 0.0in 0.2in, clip=true, width=0.5\textwidth]{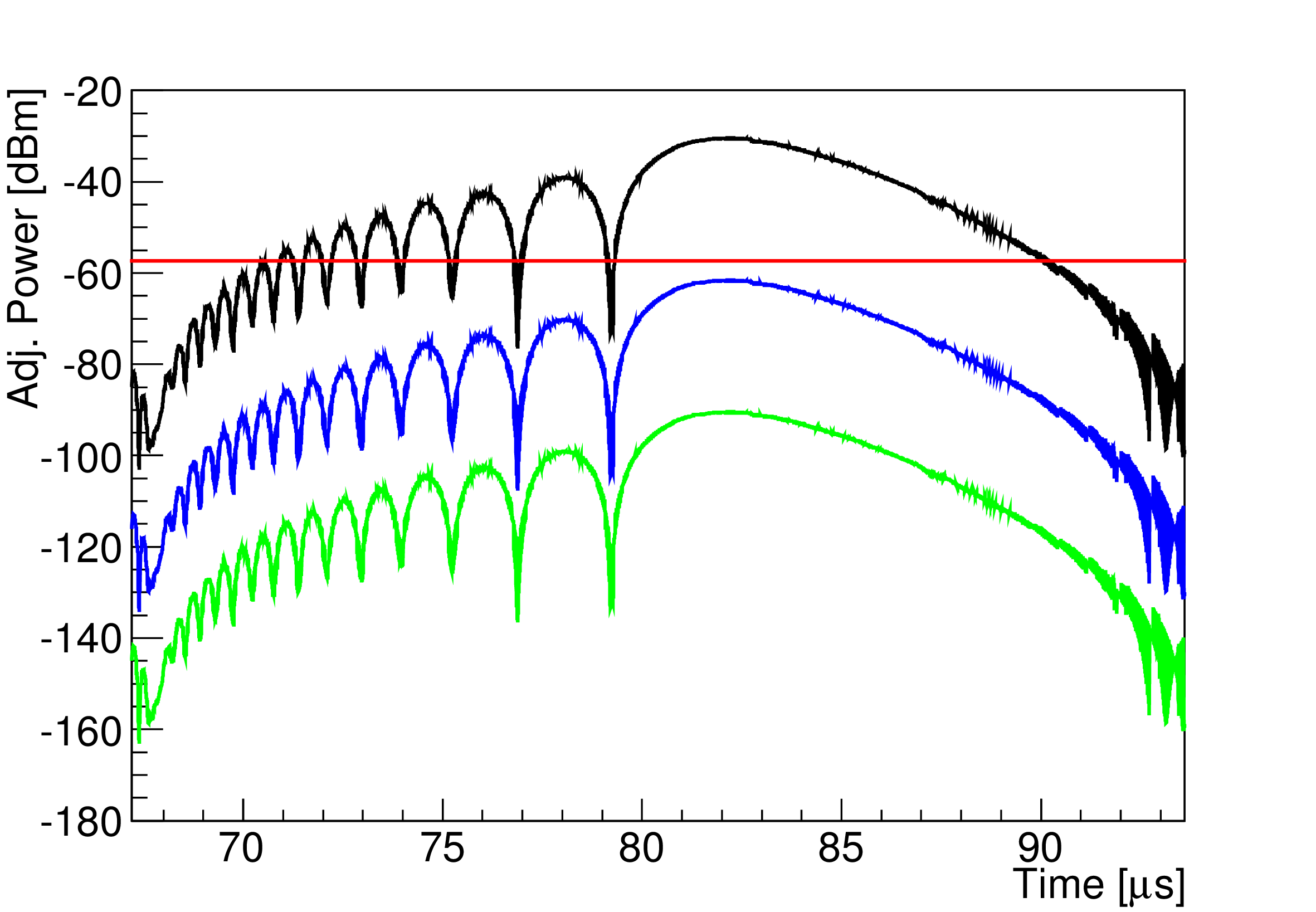}} 
{\caption{Received power vs. time for the event with $\Gamma_{90} = 0.00077$ (black) after adjusting for uncertainty in reconstruction parameters. The same received power curve is shown multiplied by $\Gamma_{90}$ (blue) or the damping factor $10^{-6}$ (green) calculated in Section~\ref{sub:plasma_scatt} to account for collisional damping. The red line is integrated background noise power in the TARA passband. 
}
\label{fig:lowsf_chirp_pow_sf90}}
\end{figure}

Figure~\ref{fig:lowsf_chirp_pow_sf90} shows simulated received power versus time for the event with $\Gamma_{90} = 0.00077$. The top (black) curve is the unmodified simulated power for this shower, within the thin-wire scattering model. The middle (blue) curve is the thin-wire simulated power for this shower modified by the $\Gamma_{90} = 0.00077$ scale factor. The bottom (green) curve is the thin-wire simulated power modified by the $10^{-6}$ collisional damping factor calculated in Section~\ref{sub:plasma_scatt}. The red line is the integrated TARA passband noise power, which is calculated from data which have been amplified by 30~dB in the RF frontend. This plot confirms earlier predictions: damping is large, and this analysis technique can detect signal below the noise floor. 

\section{Conclusion}
\label{sec:conclusion}
The TARA detector is the first facility designed specifically to search for the radar signature of UHECR. It combines a dedicated 40~kW, 54.1~MHz transmitter, high-gain (23~dB) phased antenna array, dual-polarized log periodic receiver antennas and a 250~MS/s receiver, colocated with the conventional Telescope Array cosmic ray observatory in radio-quiet western Utah, U.S.A.  

In this paper, we have reported the null result of a search for radar echoes in a set of radio waveforms which were collected by triggering on the Telescope Array Long Ridge fluorescence detector. None of the cosmic-ray coincident waveforms contained signals comparable to the high rate ``chirps'' expected for EAS radar echoes. A small subset of waveforms contained broadband transients, at a level consistent with the expectation from random snapshots of the radar receiver stream. 

Based on these observations, we have for the first time set upper limits on the radar cross sections of particular EAS. These limits are set based on the assumption that the radar cross section of EAS is dominated by a relatively narrow (few~cm) core, corresponding to the overdense regime in a plasma in the absence of collisional damping effects. Skin-depth considerations are also ignored. Such an airshower may be treated as a thin wire relative to our 5.5~meter sounding wavelength, and we designate its theoretical cross section as $\sigma_{\rm TW}$. For the most geometrically ideal EAS we studied in the present analysis, we place an upper limit of $7.7 \times 10^{-4}~\sigma_{\rm TW}$, at 90\%~c.l. 

This limit is consistent with the expectation that collisional damping of ionization electrons by neutral molecules heavily attenuates any potential radar signature in this regime of sounding frequencies and ionization densities. Further, these studies indicate that that the prospects are poor for remote detection of cosmic ray air showers via the radar technique.


\section{Acknowledgments}
\label{sec:ack} 
TARA is supported by NSF PHY-0969865, PHY-1126353 (MRI), PHY-1148091, and the W.M.~Keck Foundation. The Telescope Array experiment is supported by the Japan Society for the Promotion of Science through Grants-in-Aids for Scientific Research on Specially Promoted Research (21000002) ``Extreme Phenomena in the Universe Explored by Highest Energy Cosmic Rays'' and for Scientific Research (19104006), and the Inter-University Research Program of the Institute for Cosmic Ray Research; by the U.S. National Science Foundation awards PHY-0307098, PHY-0601915, PHY-0649681, PHY-0703893, PHY-0758342, PHY-0848320, PHY-1069280, PHY-1069286, PHY-1404495 and PHY-1404502; by the National Research Foundation of Korea (2007-0093860, R32-10130, 2012R1A1A2008381, 2013004883); by the Russian Academy of Sciences, RFBR grants 11-02-01528a and 13-02-01311a (INR), IISN project No. 4.4509.10 and Belgian Science Policy under IUAP VII/37 (ULB). The foundations of Dr. Ezekiel R. and Edna Wattis Dumke, Willard L. Eccles and the George S. and Dolores Dore Eccles all helped with generous donations. The State of Utah supported the project through its Economic Development Board, and the University of Utah through the Office of the Vice President for Research. The experimental site became available through the cooperation of the Utah School and Institutional Trust Lands Administration (SITLA), U.S. Bureau of Land Management, and the U.S. Air Force. We also wish to thank the people and the officials of Millard County, Utah for their steadfast and warm support. We gratefully acknowledge the contributions from the technical staffs of our home institutions. An allocation of computer time from the Center for High Performance Computing at the University of Utah is gratefully acknowledged.

\bibliography{refs}

\end{document}